\documentclass[fleqn,usenatbib,useAMS]{mnras}

\usepackage{graphicx}
\usepackage{amsmath}
\usepackage{amssymb}    
\usepackage{multicol}      
\usepackage{bm}
\usepackage{float}
\usepackage{chngcntr}
\usepackage{soul}
\usepackage{pdflscape}    
\usepackage{caption}
\usepackage{subcaption}
\usepackage{enumitem}

\usepackage[toc,page]{appendix}
\usepackage{xcolor}

\newcommand{\ha}[1]{H$\alpha$#1}

\newcommand{\nii}[1]{[N\,II]$\lambda$6583#1}
\newcommand{\oiii}[1]{[O\,III]$\lambda$5007#1}
\newcommand{\kms}[1]{\,km\,s$^{-1}$#1}

\title[Gas Inflows in NGC\,1097]{Composite Bulges – IV. Detecting Signatures of Gas Inflows in the IFU data: The MUSE View of Ionized Gas Kinematics in NGC\,1097}

\author[Kolcu et al.]{Tutku Kolcu$^{1, 2}$\thanks{Contact e-mail: \href{mailto:T.Kolcu@2020.ljmu.ac.uk}{T.Kolcu@2020.ljmu.ac.uk}}, Witold Maciejewski$^{1}$, Dimitri A. Gadotti$^{2,3}$, Francesca Fragkoudi$^{2,4}$, Peter Erwin$^{5,6}$, \newauthor{Patricia Sánchez-Blázquez$^{7}$, Justus Neumann$^{8}$, Glenn Van de Ven$^{8,9}$, Camila de Sá-Freitas$^{2}$}, \newauthor{Steven Longmore$^{1}$, Victor P. Debattista$^{10}$} \\
$^{1}$ Astrophysics Research Institute, Liverpool John Moores University, IC2 Liverpool Science Park, 146 Brownlow Hill, L3 5RF, UK \\
$^{2}$ European Southern Observatory, Karl-Schwartzchild-Str. 2, D-85748 Garching bei München, Germany\\
$^{3}$ Centre for Extragalactic Astronomy, Department of Physics, Durham University, South Road, Durham DH1 3LE, UK \\
$^{4}$ Institute for Computational Cosmology,  Department of Physics, Durham University, DH1 3LE, United Kingdom\\
$^{5}$ Max-Planck-Insitut für extraterrestrische Physik, Giessenbachstrasse, 85748 Garching, Germany \\
$^{6}$ Universitäts-Sternwarte München, Scheinerstrasse 1, D-81679 München, Germany \\
$^{7}$ Departamento de Física Teórica, Universidad Autónoma de Madrid, E-28049 Cantoblanco, Spain \\
$^{8}$ Max-Planck Institut für Astronomie, Königstuhl 17,
D-69117, Heidelberg, Germany \\
$^{9}$ Department of Astrophysics, University of Vienna, Türkenschanzstrasse 17, 1180 Vienna, Austria \\
$^{10}$ Jeremiah Horrocks Institute, University of Central Lancashire, Preston, PR1 2HE, UK \\ }

\date{Accepted XXX. Received YYY; in original form ZZZ}

\begin{document}

\maketitle

\begin{abstract}
Using VLT/MUSE integral-field spectroscopic data for the barred spiral galaxy NGC\,1097, we explore techniques that can be used to search for extended coherent shocks that can drive gas inflows in centres of galaxies. Such shocks should appear as coherent velocity jumps in gas kinematic maps, but this appearance can be distorted by inaccurate extraction of the velocity values and dominated by the global rotational flow and local perturbations like stellar outflows. We include multiple components in the emission-line fits, which corrects the extracted velocity values and reveals emission associated with AGN outflows. We show that removal of the global rotational flow by subtracting the circular velocity of a fitted flat disk can produce artefacts that obscure signatures of the shocks in the residual velocities if the inner part of the disk is warped or if gas is moving around the centre on elongated (non-circular) trajectories. As an alternative, we propose a model-independent method which examines differences in the LOSVD moments of \ha~and \nii. This new method successfully reveals the presence of continuous shocks in the regions inward from the nuclear ring of NGC\,1097, in agreement with nuclear spiral models.
\end{abstract}

\section{Introduction}
Gas inflows play an important role in the evolution of galaxies: they enhance nuclear star formation (SF) \citep{Ellison_11,Prieto_19}, which can cause starbursts in nuclear rings \citep{Heller_1994, Mazzuca_08} and change the morphology of the central regions by creating substructures (e.g., gaseous and stellar disks, rings) \citep{Shlosman_1989, Athanassoula_1992, Martini_03, Kim_2012,Sormani_15}. Further, the gas reaching the very centre can trigger active galactic nuclei (AGN) \citep{Davies_14, Audibert-Combes-2020} and lead to highly energetic feedback \citep{Prieto_05, Davies_14, Audibert_19}. Outflowing gas from AGN feedback can in turn rapidly suppress SF and cause quenching \citep{Man_19, Donnari_21}. Hence, to understand the structural evolution of the galaxies, it is important to understand the role of the inflowing gas.
 
Mechanisms triggering gas inflow on various scales have been a prominent focus of many observational and theoretical studies \citep{Sanders_1976, Athanassoula_1992, Emsellem_03, Maciejewski_2004, Fathi_06, Lin_13, Fragkoudi_16,Prieto_19}. Simulations have revealed that non-axisymmetric potentials such as large-scale bars and ovals efficiently transport gas towards the inner kiloparsec \citep{Athanassoula_1992, Garcia-Burillo-05, Lin_13, Audibert_21} by generating torques that put gas clouds on intersecting trajectories. This leads to shocks, in which the gas loses angular momentum and funnels inwards, though the efficiency of this mechanism inside the inner kiloparsec is questioned \citep{Kim_2012}. It is also evident from both observations and simulations that central stellar components such as nuclear bars and disks affect the kinematics of the gas \citep{Shlosman_01,Maciejewski_02, Fathi_05}, but we require more understanding of mechanisms transferring gas in the nuclear regions of the galaxies.

Shocks in gas that form in response to the gravitational torques of bars and ovals extend over large distances along straight or curved lines, the latter often assuming a spiral shape. Although dust filaments can indicate gas compression in shocks, shock condition implies a rapid change in velocity \citep{Athanassoula_1992,Maciejewski_2004}, therefore coherent structures in kinematic maps are best indicators of shocks \citep{Zurita_04, Fathi_05, Storchi-Bergmann-07}. In this series of papers, we aim to identify gas inflows by searching for coherent kinematic structures in integral-field spectroscopic observations, which allow us to obtain spectral information over a two-dimensional spatial field-of-view (FoV) of integral field units (IFUs).  

We will conduct our search using a diverse sample of galaxies from a multi-wavelength imaging and spectroscopic observing campaign \textit{Composite Bulges Survey}  \citep[\textit{CBS},][]{Erwin_2023}. CBS focuses on understanding the formation and growth mechanisms of bulges, and the correlation with their host galaxy properties and aims to identify and characterise different forms of nuclear stellar components, from large-scale bars down to nuclear scale discs/rings/bars and nuclear star clusters, of nearby galaxies \citep{Erwin_21}. The sample is a mass- and volume-limited set of 53 disk (S0--Sbc) galaxies with optical and near-IR (NIR) imaging with the Hubble Space Telescope (HST), of which 40 have been observed with the Multi Unit Spectroscopic Explorer (MUSE) instrument of the Very Large Telescope (VLT). Therefore, benefiting from a diverse galaxy sample, we should be able to determine whether gas inflows in the centres of galaxies occur predominantly in coherent kinematic structures. Moreover, we should be able to examine the connection between different central stellar structures and the presence of coherent kinematic structures indicating inflow.

Tracing the flow of gas in nuclear regions of galaxies has been a challenging task with former generations of IFUs because of their instrumental limitations, such as small fields of view and low spatial resolution. Moreover, in the case of having a low signal-to-noise ratio (SNR), coherent kinematic structures were lost after binning \citep{Fathi_05, Fathi_06, Storchi-Bergmann-07}. In our study, we predominantly rely on data from MUSE, which with its large, 60\arcsec$\times$60\arcsec, FoV, fine spatial resolution and 0.2\arcsec$\times$0.2\arcsec~sampling, allows us to capture the nuclear regions of nearby galaxies in great detail. Moreover, MUSE can obtain sufficient SNR for nearby targets, minimising the need for binning to improve SNR. For our study on gaseous kinematic structures, we prefer not to bin the data to prevent losing spatial information. Hence, in comparison to pioneering studies that suffered from instrumental limitations, advancements of MUSE maximise the prospect of finding coherent kinematic structures.

In this pilot study, we introduce the methodology that we developed to identify coherent kinematic structures in the centres of galaxies, and apply it to the nuclear regions of a galaxy in the CBS sample, NGC\,1097 \citep[D=14.8 Mpc,][]{Freedman_2001}, 1\arcsec\,$\approx$\,72\,pc). NGC\,1097 is an SBb galaxy with a strong bar at position angle $\sim$141\degr \citep{Prieto_05} and it contains a prominent star-forming nuclear ring in the inner $\sim$8\arcsec--10\arcsec  \citep{Bittner_20}. The galaxy hosts a LINER\footnote{LINER: Low Ionization Narrow Emission-line Regions, can be distinguished from the Seyfert type nuclei based on their [O\,III] luminosity and kinematics \citep{Kewley_06}} type, low-luminosity active galactic nucleus \citep[LLAGN -- $L_{bol}\lesssim10^{42}\,erg\,s^{-1}$,][]{Ho_08} with bolometric luminosity $L_{bol}=8.6 \times 10^{41}\,erg\,s^{-1}$ \citep{Nemmen_06}. The galaxy is interacting with a companion elliptical galaxy NGC\,1097A located
to its northwest. Because of its complex characteristics, NGC\,1097 has been a prominent focus of many studies \citep{Quillen_95, Prieto_05, Fathi_06,  Davies_09,vandeVen_Fathi_2010,Lin_13, Bowen_16, Izumi_17, Prieto_19, Legodi_21}.

The flow of the paper is as follows: In Section \ref{sec:data} we present the observations of NGC\,1097, and we introduce the data extraction processes for MUSE, including our approach of fitting the gas emission with multiple components. We present our main methodology of finding coherent kinematic structures, and we apply it to the galaxy NGC\,1097 in Sect. \ref{sec:method}. We provide the discussion of our findings in Sect. \ref{sec:discussion}.

\section{Observations and Procedures for deriving the LOSVD moments}
\label{sec:data}

\subsection{Observations of NGC\,1097}
\label{sec:MUSE-Sample}
\begin{figure*}    
    \hspace{-1cm}
    \includegraphics[width=1.7\columnwidth]{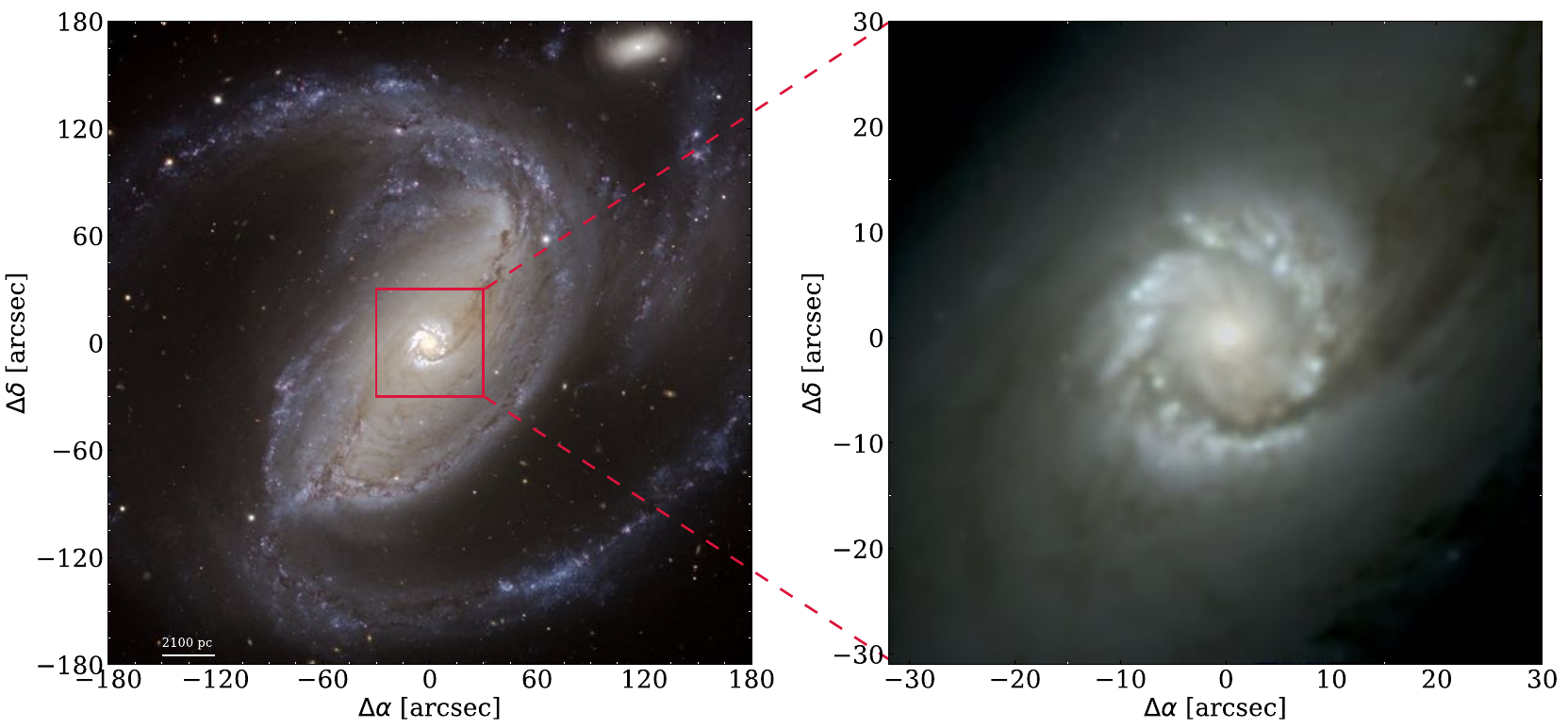}
    \caption{MUSE coverage of NGC\,1097. \textit{left:} Large-scale (360\arcsec$\times$360\arcsec) image of NGC\,1097 taken by VIMOS/VLT. The MUSE coverage of 60\arcsec$\times$60\arcsec~is highlighted with the red box. \textit{right:} Colour composite image within the MUSE FoV, constructed by using the MUSE data cube. Images in both panels are centred on the brightest point in the centre. The North is to the top and the East is to the left.}
    \label{fig:1097_vimos_cflux}
\end{figure*}
MUSE observations of NGC\,1097 used in CBS are provided by the \textbf{TIMER} project as a part of observing proposal ID:097.B-0640 (PI: Dimitri Gadotti), carried out in ESO Period 97 on 31$^{st}$ of July 2016. The TIMER project studies nearby barred galaxies ($D<40Mpc$) larger than 1 arcmin with stellar masses above $10^{10}M_{\odot}$, brightness threshold of 15.5 B-mag and inclinations smaller than $\approx$60\degr, which display  prominent central components (e.g., nuclear rings/discs, inner bars). Observations are done in MUSE Wide-Field Mode (WFM), for a nominal wavelength range of $4700-9350$\AA. The FoV of WFM encloses an area of 60\arcsec$\times$60\arcsec, has a spatial sampling of 0.2\arcsec/pixel \citep{Bacon_10}. The point spread function full width at half maximum of the observations for NGC\,1097 is 0.8\arcsec, which corresponds to a spatial resolution of $\sim$58 pc. Further details on observations and the data reduction processes, products including stellar kinematics, ages and metallicities of the sample can be found in \citet{Gadotti_19,Gadotti_20} and \citet{Bittner_20}. In the left panel Fig. \ref{fig:1097_vimos_cflux}, we present a large-scale (360\arcsec$\times$360\arcsec) image\footnote{Image credits: \citealp{PR_release_NGC1097}, date of observation: 09-10/12/2004, image ID: eso0438d, media source: \url{https://www.eso.org/public/images/eso0438d/}} of NGC\,1097, which reveals the spiral arms on the north and the south, and the companion elliptical galaxy NGC\,1097A in the northwest. In the right panel Fig. \ref{fig:1097_vimos_cflux}, we present the colour composite image within the inner 60\arcsec$\times$60\arcsec, constructed from the MUSE data. In both panels, the stellar nuclear ring in the centre and the connection of dust lanes to the nuclear ring are very well resolved.

\subsection{Extraction of the kinematic data}
\label{sec:data_extraction}
\begin{figure}
\centering
    \hspace{-0.25cm}
    \includegraphics[width=0.5\columnwidth]{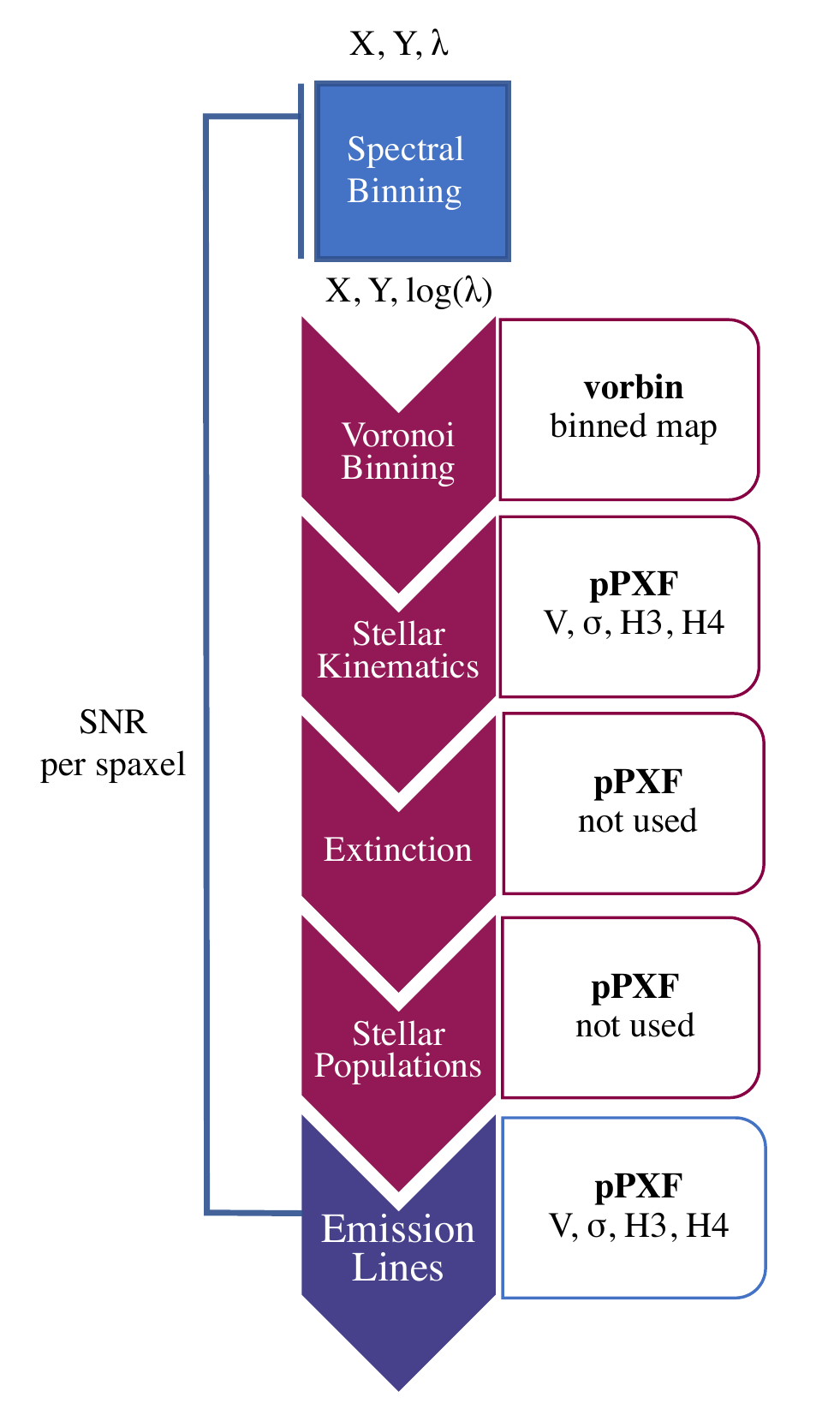}
    \caption{Our implementation of the Data Analysis Pipeline. Arrows follow the order of the workflow within the pipeline. Voronoi binned data is transferred to stellar kinematics, extinction (not used), stellar population (not used) and emission line analysis modules respectively. Note that the emission line analysis can also be performed on single spaxels. In the rightmost blocks, we present the routine used in each module to perform the analysis alongside the derived products.}
    \label{fig:DAP}
\end{figure}

Spatially resolved spectroscopic data hold a great amount of information, which can be used to probe the fundamental properties of galaxies. 
To derive emission line fluxes and to extract stellar and ionised gas kinematics from MUSE data, we employ the Data Analysis Pipeline\footnote{\texttt{DAP} can be accessed at: \url{https://gitlab.com/francbelf/ifu-pipeline}} (\texttt{DAP}) of the PHANGS collaboration \citep{Emsellem_22}, which is an implementation based on the GIST pipeline \citep{Bitner_19}. \texttt{DAP} is an exceptionally functional software of modular structure, and it employs the commonly used penalised pixel-fitting (\texttt{pPXF}) code \citep{Cappellari-emsellem-2004} to perform both stellar and emission line analysis. To optimize the processing time of \texttt{DAP}, we discarded the stellar population history module, since it was outside of the scope of this work. A schematic overview of \texttt{DAP} highlighting the main modules, the routines for data extraction and the products is presented in Fig. \ref{fig:DAP}. A representative spectrum with the \texttt{pPXF} fit for both stellar and emission lines is shown in Fig. \ref{fig:representative_excess_spectra} (top panel). 

\subsubsection{Deriving stellar line-of-sight velocity distribution moments}
\label{sec:stellar_extraction}

We extracted the stellar kinematics in the form of line-of-sight velocity distribution (LOSVD) moments \textemdash \  mean velocity (V), velocity dispersion ($\sigma$) and Gauss-Hermite higher-order moments ($h_3$ \& $h_4$) \citep{van_der_Marel_1993}. Before deriving the LOSVD moments, the data is re-sampled on a logarithmic wavelength axis, then it is spatially binned with adaptive weighted implementation \citep{Diehl_06} of the Voronoi tessellation \citep{Cappellari_Copin_03}. For the spectral fitting, we limited the rest frame wavelength range to 4800-7000\AA. To fit the stellar continuum with \texttt{pPXF}, we used the EMILES single-stellar population (SSP) model library \citep{Vazdekis_15}.  SSP models convolved with the LOSVD are fitted to each spectrum and the best-fitting parameters are determined by $\chi^2$ minimisation in pixel space. We used a threshold SNR of 3 for spaxels included in Voronoi bins, to prevent contribution from low-quality spaxels whose signal is below the isophote level. We also used a target SNR of 40 for each Voronoi bin, as this value is commonly used in literature and provides a balance between the accuracy of the extracted measurements and the spatial resolution \citep{van_der_Marel_1993}.

When fitting the stellar kinematics, we used 8$^{th}$ order multiplicative Legendre polynomials within the \texttt{pPXF} routine and included no additive polynomials. Multiplicative polynomials adjust the inaccuracies in the spectral calibration and make the fits independent of reddening by dust so that we do not need to provide a particular reddening curve within the routine \citep{Cappellari_2017}. Although additive polynomials can minimise mismatches between stellar templates and absorption lines and correct impurities from the sky subtraction \citep{Cappellari_2017}, they can also alter the equivalent width of the Balmer absorption lines and might introduce non-physical correction to the line fluxes \citep{Emsellem_22}. Therefore, we proceeded to use only multiplicative polynomials, both to avoid any non-physical flux values and for the efficiency of the routine since combining both additive and multiplicative Legendre polynomials is computationally expensive. 

\begin{figure}
    \centering
    \includegraphics[width=1\columnwidth]{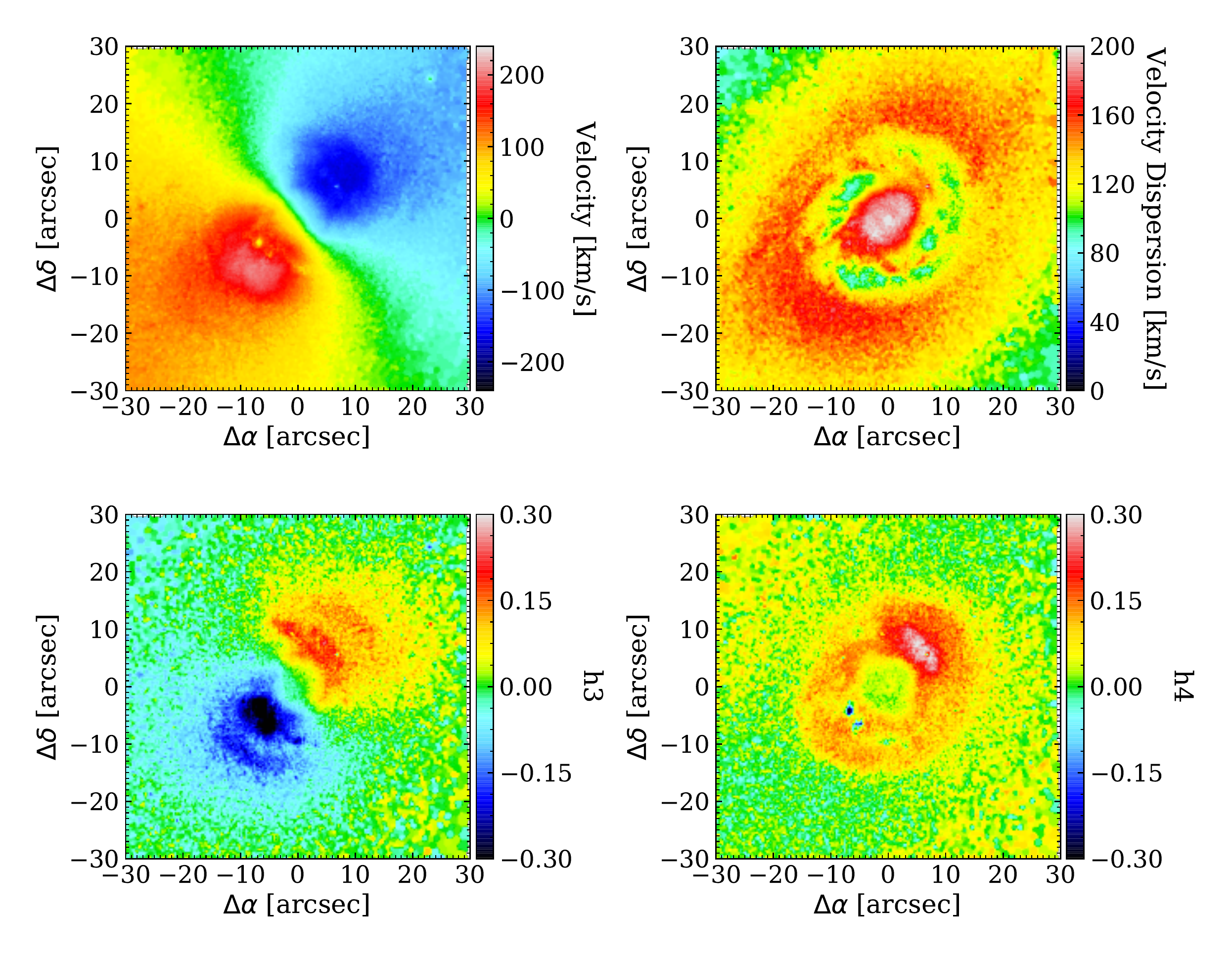}
    \caption{Stellar LOSVD moment maps of NGC\,1097. Stellar velocity (top left) velocity dispersion (top right) Gauss-Hermite moments h3 (bottom left) and h4 (bottom right)}
    \label{figapp:stellarmaps}
\end{figure}

The stellar LOSVD moments can be seen in Fig. \ref{figapp:stellarmaps}. For the systemic velocity of NGC\,1097, we used $1271$ \kms, taken from the NASA/IPAC Extragalactic Database (NED\footnote{NED can be accessed via \url{https://ned.ipac.caltech.edu/}}) and based on the redshift measurement in \citealp{Allison_14} ($z$\,$\approx$\,0.00424). Further in this work, all wavelengths are corrected for the redshift of the galaxy, equivalent to the rest frame wavelengths of the lines.

To define the centre of the galaxy, we first attempted to fit ellipses to the flux distribution by using the \textit{ellipse} routine of IRAF\footnote{IRAF: the Image Reduction and Analysis Facility software \citep{IRAF_2000}} software, but we realised that the flux distribution of the galaxy is asymmetric, so the parameters of the ellipses were poorly constrained. Hence, we defined the photometric centre as the brightest spaxel in the nucleus based on the flux integrated over the entire MUSE spectral range.
The kinematic centre derived in Sect. \ref{sec:method-vres_emissionlines} differs only by $\sim$3 spaxels from thus defined photometric centre. Hereafter, we give the coordinates $\Delta\alpha$, $\Delta\delta$ in relation to the photometric centre, i.e. the photometric centre is located at [$\Delta\alpha$,$\Delta\delta$]=[0,0].

\subsubsection{Deriving ionised gas line-of-sight velocity distribution moments}
\label{sec:gas-kin-extraction}
\begin{table}
\centering
\caption{Characteristics of the fitted emission lines. Wavelengths of the emission lines are taken from the National Institute of Standards and Technology, (NIST; \url{https://www.nist.gov/pml/atomic-spectra-database}).}
\begin{tabular}{ccc}
\hline
Line ID & \begin{tabular}[c]{@{}c@{}}Wavelength \\ (\AA)\end{tabular} &  \begin{tabular}[c]{@{}c@{}}Amplitude Ratio\\ (fixed)\end{tabular} \\ \hline
\multicolumn{3}{c}{Group 1 - Hydrogen Balmer Lines} \\ \hline
H$\beta$ & \hspace{0.5cm} 4861.35  &  \\
H$\alpha$ & \hspace{0.5cm} 6562.79  &  \\ \hline
\multicolumn{3}{c}{Group 2 - Low Ionisation Lines} \\ \hline
{[}N\,I{]}$\lambda$5198 &\hspace{0.5cm}  5197.90  &  \\
{[}N\,I{]}$\lambda$5200 & \hspace{0.5cm} 5200.26 &  \\
{[}N\,II{]}$\lambda$5754 &\hspace{0.5cm}  5754.49  &  \\
{[}O\,I{]}$\lambda$6300 &\hspace{0.5cm}  6300.30  &  \\
{[}O\,I{]}$\lambda$6364 &\hspace{0.5cm}  6363.78  & \hspace{0.5cm} 0.33[O\,I]$\lambda$6300 \\
{[}N\,II{]}$\lambda$6548 & \hspace{0.5cm} 6548.05  & \hspace{0.5cm} 0.34{[}N\,II{]}$\lambda$6583 \\
{[}N\,II{]}$\lambda$6583 &\hspace{0.5cm}  6583.45  &  \\
{[}S\,II{]}$\lambda$6716 &\hspace{0.5cm}  6716.44  &  \\
{[}S\,II{]}$\lambda$6731 &\hspace{0.5cm}  6730.82  &  \\ \hline
\multicolumn{3}{c}{Group 3 - High Ionisation Lines} \\ \hline
{[}O\,III{]}$\lambda$4959 & \hspace{0.5cm} 4958.91 & \hspace{0.5cm} 0.35{[}O\,III{]}$\lambda$5007 \\
{[}O\,III{]}$\lambda$5007 & \hspace{0.5cm} 5006.84  &  \\
He\,I $\lambda$5876 & \hspace{0.5cm} 5875.61  &  \\
{[}S\,III{]}$\lambda$6312 & \hspace{0.5cm} 6312.06  &  \\
\hline

\end{tabular}
\label{tab:tracers}
\end{table}

\begin{figure*}
    \centering
    \includegraphics[width=0.65\columnwidth]{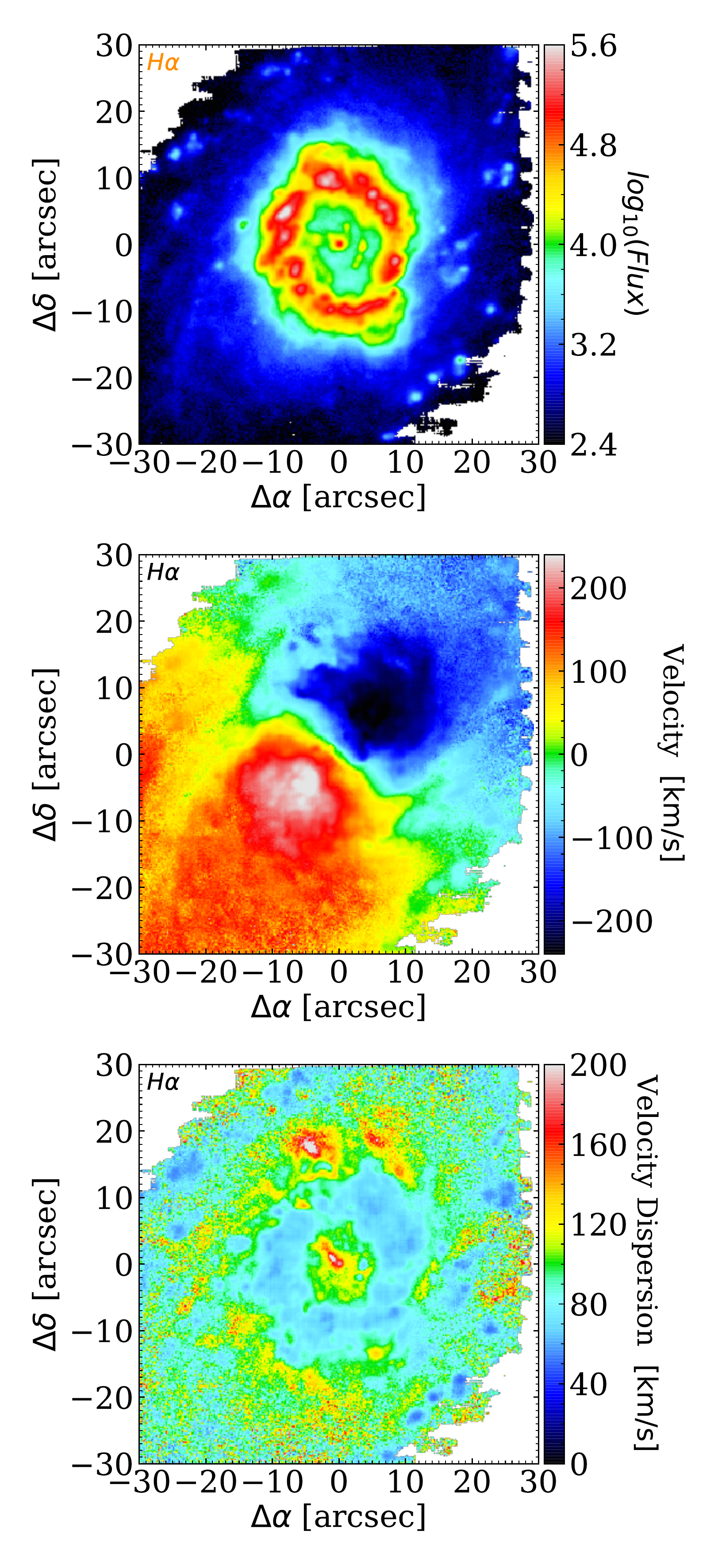}
    \includegraphics[width=0.65\columnwidth]{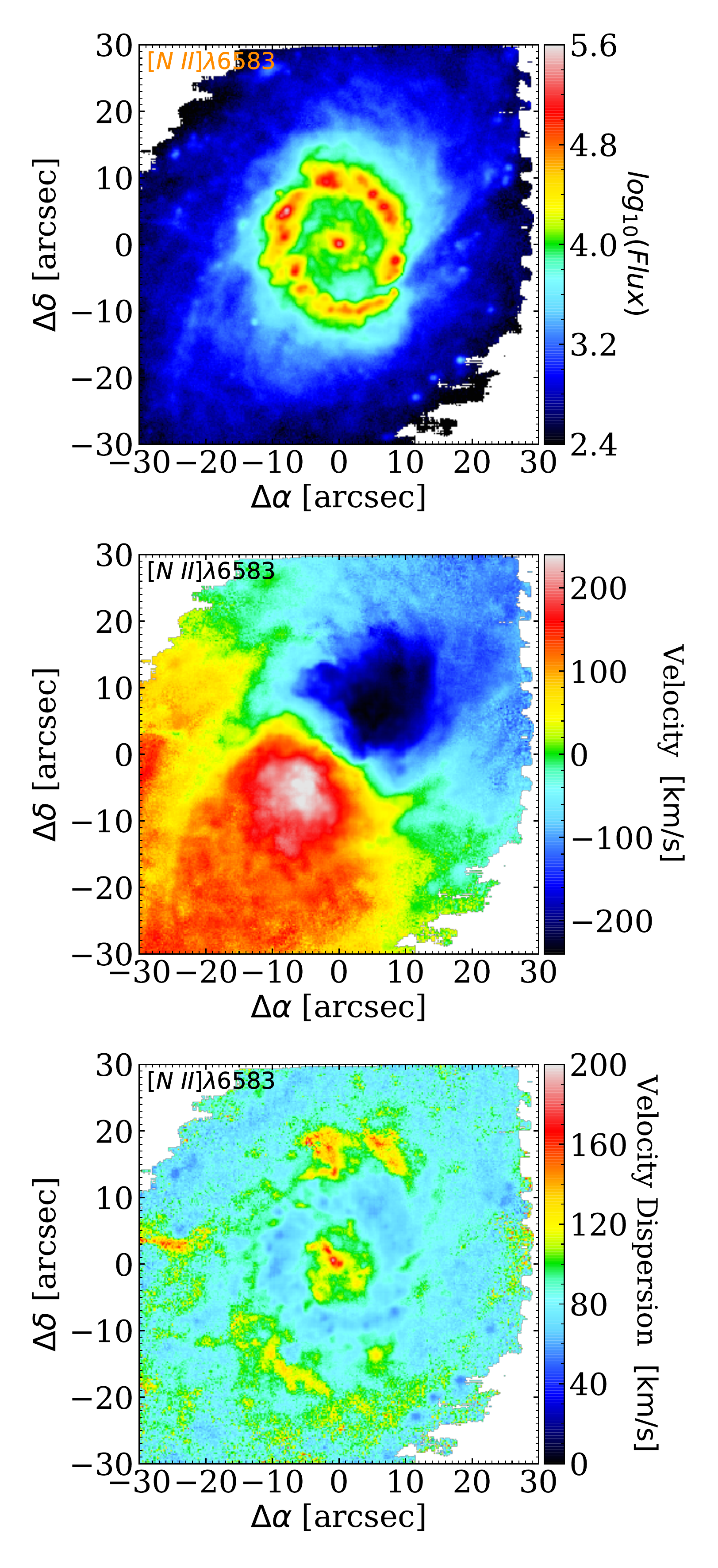}
    \includegraphics[width=0.65\columnwidth]{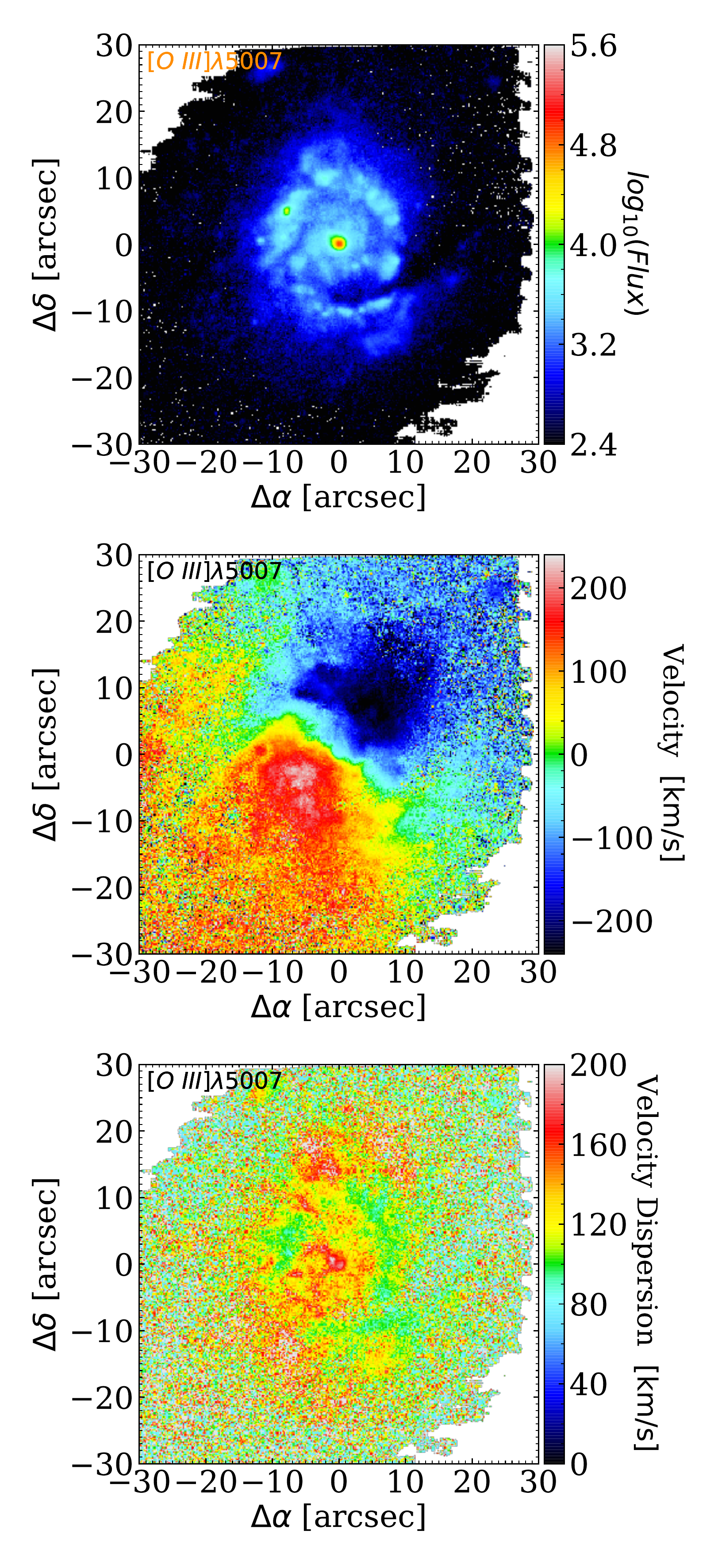}

 \caption{Flux and LOSVD moment maps of H$\alpha$ (left) \nii\AA~(middle), \oiii\AA~(right). We show the flux on logarithmic scale in units $10^{-20}$ergs$^{-1}$cm$^{-2}$spaxel$^{-1}$ (top row) velocity (middle row) and the velocity dispersion (bottom row).}

\label{fig:HA_NII_OIII_gaskinematics}
\end{figure*}

We derived the ionised gas kinematics by fitting initially single Gaussian components to the emission lines, with the stellar continuum and absorption lines fitted simultaneously. The fits were performed on individual spaxels, while the stellar velocity moments were fixed to the values determined for each Voronoi bin, specified in the prior stellar kinematics fits. Computing Gauss-Hermite higher-order moments is also possible for the emission line analysis, and we show the results of the fit including the h3 and h4 coefficients in Appendix \ref{app_supplementary_maps} (Fig. \ref{fig:NGC1097_DAP_h3h4}). However, we find fitting emission lines with Gaussians to be a better approach when searching for multiple components (see Sect. \ref{sec:excess}), whose presence otherwise could be incorporated in the higher-order moments.  

\texttt{DAP} can fit each emission line separately or fit several emission lines together by assigning each the same velocity and velocity dispersion and by imposing flux ratios. We left the fluxes free, except for the constraints from atomic physics \citep{Osterbrock_book}. We gathered the emission lines into three main groups, with kinematics shared among lines in each group. Extracting kinematics for each group separately allows us to study and understand how gas regions of different physical properties move with respect to each other. The grouping of the emission lines is done as follows: Group 1 consists of Balmer lines \textemdash H$\beta$$\lambda$4861, H$\alpha$$\lambda$6562 (indexed as H$\beta$ and \ha~in figures); Group 2 consists of low-ionisation lines \textemdash [N I]$\lambda\lambda$5197,5200, [N\,II]$\lambda$5754, [O I]$\lambda\lambda$6300,64, [N\,II]$\lambda\lambda$6548,83 ,  [S II]$\lambda\lambda$6717,31 and Group 3 consists of high-ionisation lines \textemdash He I$\lambda$5875, [O\,III]$\lambda\lambda$4959,5007, [S III]$\lambda$6312. We summarise the characteristics of the fitted emission lines in each group in Table \ref{tab:tracers}.

For each group, in Figure \ref{fig:HA_NII_OIII_gaskinematics}, we present their kinematics accompanied by the flux maps of the strongest lines in each group, which are \ha, \nii~and \oiii~respectively. To avoid contribution from spaxels dominated by noise, we used a threshold SNR$=$10; spaxels of lower SNR are discarded from the maps. All three groups show high flux in the nuclear ring and in the nucleus. The observed velocity field of each group has a typical spider velocity pattern and is dominated by rotation, but it also shows strong distortions from circular motion caused by the presence of the bar. Such distortions in the gas kinematics have been reported in various nearby galaxy studies \citep{Bosma_1978,Emsellem_01, Zurita_04, Fathi_04,Fathi_06}. The velocity dispersion is the lowest in the nuclear ring and in the outer regions and significantly increases close to the nucleus.

\subsection{Improving the quality of the fits by fitting multiple Gaussians}
\label{sec:excess}
\begin{figure}
    \centering
    \includegraphics[width=1\columnwidth]{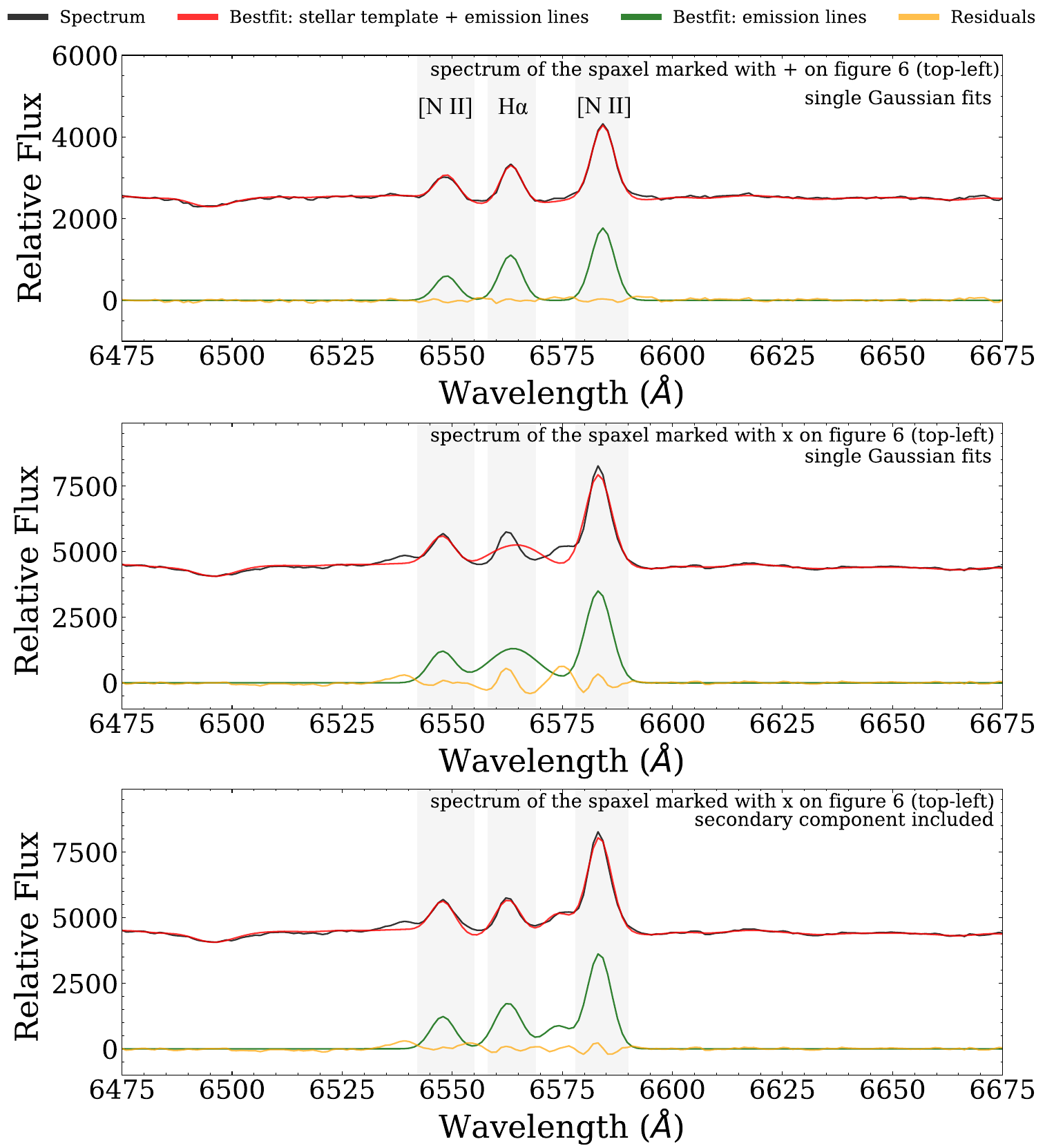}
    \caption{Representative spectra zoomed on the region of \ha~and \nii, and the best fits from \texttt{pPXF}/\texttt{DAP}. The lines represent the MUSE spectrum (black), the combined best fit of the stellar template and emission lines (red), the best fit of emission lines (green), and the difference between the spectrum and the combined best-fit stellar template with emission lines (orange). \textit{top:} Spectrum of a spaxel without the excess, marked in the upper left panel of Fig. \ref{fig:vel-sigma-R12R26} with a '+'. \textit{middle:} Spectrum of a spaxel with the excess affecting the single Gaussian fits of \ha~and \nii, marked in the upper left panel of Fig. \ref{fig:vel-sigma-R12R26} with an 'x'. \textit{bottom:} Spectrum of the spaxel shown in the middle panel, with fits that include the secondary component.}
    \label{fig:representative_excess_spectra}
\end{figure}
\begin{figure}
\centering
    \includegraphics[width=1\columnwidth]{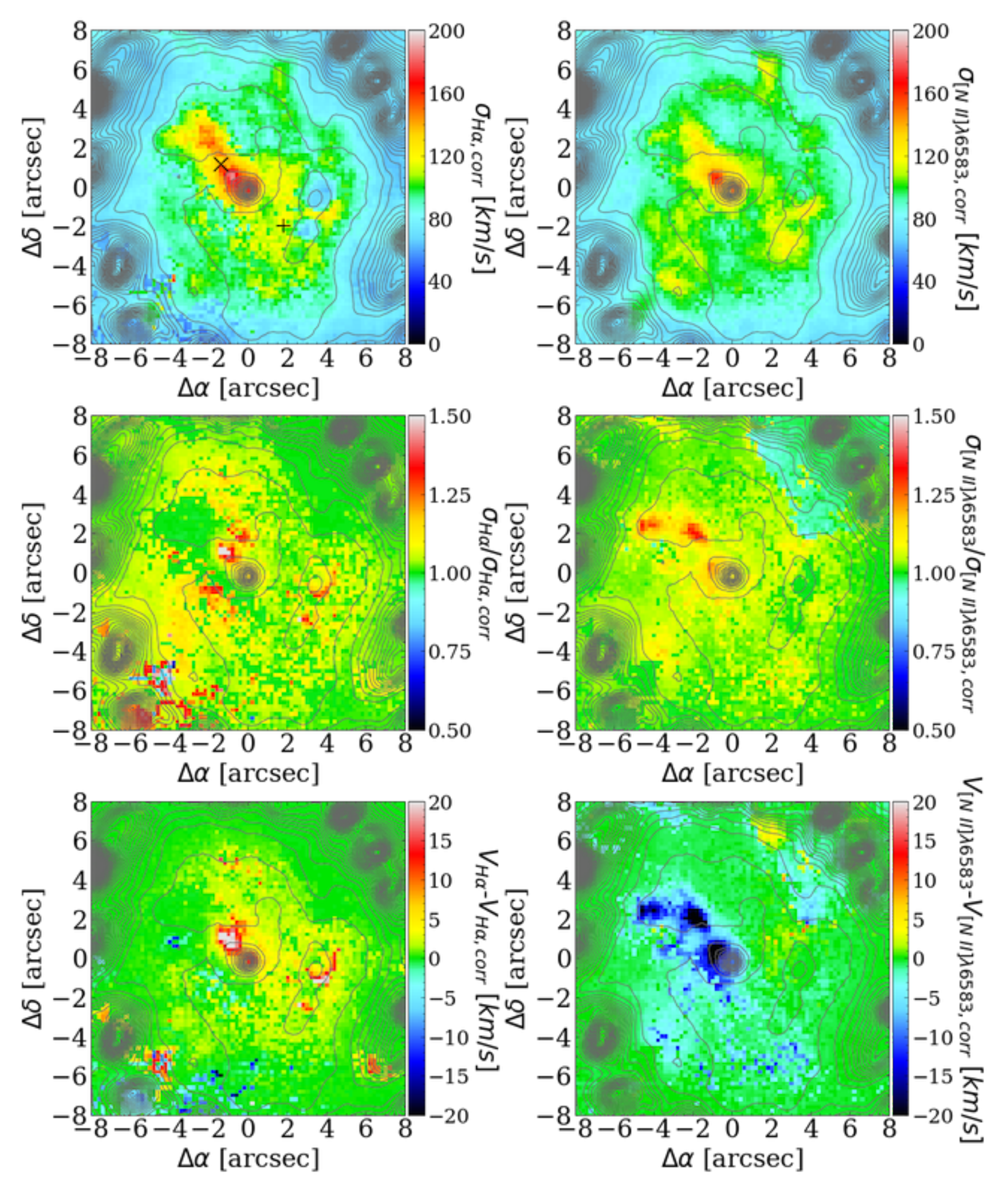}
    \caption{Comparison between fits using a single Gaussian and fits including a secondary \nii~component. Grey contours in all panels highlight the \ha~emission. The left and right columns represent \ha~and \nii~respectively. \textit{top:} the corrected velocity dispersion. 'x' and '+' markers on the left panel represent the spaxels whose spectra are displayed in Fig. \ref{fig:representative_excess_spectra}. \textit{middle:} The ratio of velocity dispersion obtained from single Gaussian fits to the corrected velocity dispersion. \textit{bottom:} The velocity differences between the corrected velocity field and the velocity field obtained with single Gaussians. }
    \label{fig:vel-sigma-R12R26}
\end{figure}

\begin{figure*}
\centering
\includegraphics[width=2.1\columnwidth]{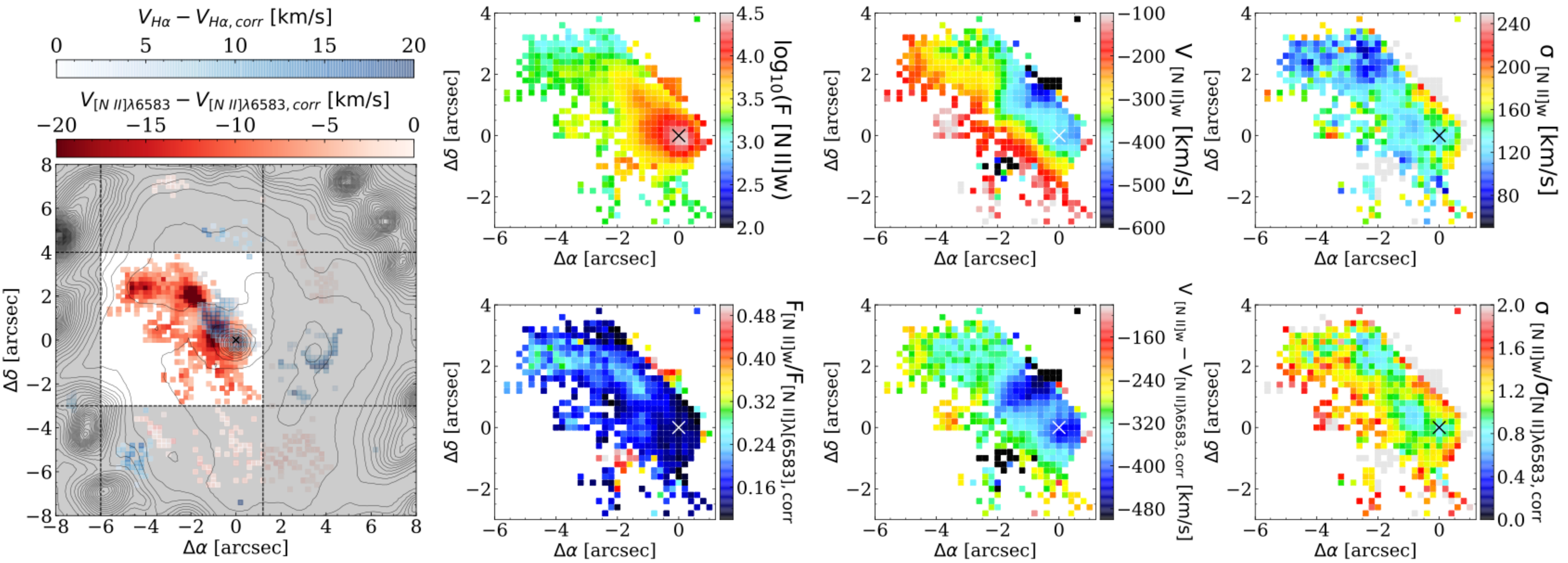}
    \caption{\textit{left:} The mask of spaxels for which the fits including secondary component were used in further analysis. Spaxels in the grey-shaded region fall outside the defined conditions and are excluded from the mask. Blue and red colours, cut respectively at the 5\kms~and 4\kms~thresholds, show the difference between velocities extracted by fitting single Gaussians and after including the secondary component for \ha~and \nii. Grey contours represent the \ha~emission. The centre of the galaxy is marked with a cross. \textit{right:} Flux and kinematics of the secondary component within the mask. Only spaxels within the mask are shown. Maps respectively show the derived flux (top left), velocity (top middle), velocity dispersion (top right), the ratio of the fluxes of the secondary component and the main \nii~line (bottom left), the difference in velocities of the secondary component and the main \nii~line (bottom middle), the ratio of velocity dispersion of the secondary component and the main \nii~line (bottom right).}
    \label{fig:mask_kinematics_excess}
\end{figure*}
In the regions inner to the nuclear ring, the kinematics of all emission line groups reveal a region of high velocity dispersion, extending from the centre of the galaxy $\sim$3\arcsec~towards the northeast (see Fig. \ref{fig:HA_NII_OIII_gaskinematics}), which drove us to closely examine the spectra in the affected spaxels. In Figure \ref{fig:representative_excess_spectra}, we show the \ha~and [N\,II]$\lambda\lambda$6548,83 region of the spectrum in one of the affected spaxels, as well as in a spaxel outside of the affected region. These spaxels are located at the positions marked by an 'x' and a '+', respectively, in the top left panel of Fig. \ref{fig:vel-sigma-R12R26}. In the spectrum of the affected spaxel, as shown in the middle panel of Fig. \ref{fig:representative_excess_spectra}, we notice excess emission in the blue wing of \nii~line, with a peak at the rest frame wavelength $\sim$6574.50\AA, which hereafter we call the \textit{excess}. Because of the excess, the line profile of \nii~becomes strongly non-Gaussian, and the excess encroaches on the region between \nii~and \ha, so the \ha~fits are also strongly affected by its presence. As we demonstrate in the top panel of Fig. \ref{fig:representative_excess_spectra}, when the excess is not present the emission lines are well-fitted, but when the excess is present the minimisation algorithm of \texttt{pPXF} produces a broad Gaussian, especially for \ha, leading to overestimated velocity dispersion values. The presence of the excess shifts the centre of the fitted Gaussians for the main \nii~component to slightly lower velocities, and for the \ha~component to higher velocities. We found no signs of an excess in the red wing of \nii, or either of the wings of \ha. Although we see a weak excess in the blue wing of [N\,II]$\lambda$6548, its SNR is at the same level as noise, and it is insignificant compared to the excess in the blue wing of \nii. Moreover, we found no signs of an excess in the wings of the \oiii, though its presence might be lost to the noise. However, in the spaxels of the region with high velocity dispersion, where the presence of the excess in the blue wing of \nii~is the most pronounced, \oiii~line is well-fitted. This suggests that the high velocity dispersion of \oiii~in those regions (Fig. \ref{fig:HA_NII_OIII_gaskinematics}) is intrinsic.

In order to account for the excess and to improve the line fits, we introduced one additional emission component to the \texttt{DAP} configuration at wavelength $\lambda\sim$6574.50\AA, which we call the \textit{secondary component}. Throughout the rest of this work, we work under the assumption that the secondary component is a Doppler-shifted component of \nii. We tested how the derived kinematics depends on the initial guess of the wavelength of the secondary component by shifting it by $\pm$3.3\AA~(roughly equal to the dispersion broadening); we find no significant changes to the results. In the bottom panel of Fig. \ref{fig:representative_excess_spectra}, we demonstrate that for the spaxel, whose spectrum and single Gaussian fits were affected by the presence of the excess, the addition of the secondary component accounts for the excess and fixes the erroneous fits. However, one must be careful, as for the spaxels where the excess is not strong the fits of the secondary component would be poorly constrained, and we must carefully determine the spaxels truly affected by the presence of the excess. This can be done by comparing the measurements extracted from fits with and without the secondary component. As we explained earlier in this section, the excess emission on one side of the line not only increases the velocity dispersion of a single Gaussian fit but also alters its velocity. While high velocity dispersion can be intrinsic, a large change in the derived velocity caused by including the secondary component indicates strong asymmetry of the line profile, hence, likely the presence of multiple line components. Therefore, we search for the spaxels that are most affected by the presence of the excess by looking for a change in derived velocities, which should be more reliable than using the high velocity dispersion measurements.

In order to identify which spaxels in and around the region of the elevated velocity dispersion in the nucleus are affected by the presence of the excess, we fitted the secondary component in the inner 8\arcsec$\times$8\arcsec. In Figure \ref{fig:vel-sigma-R12R26}, we show the change in velocity (V) and velocity dispersion ($\sigma$) between the original values obtained from single Gaussian fits and the corrected values obtained from fits with the secondary component included. For the corrected measurements, in addition to the respective emission line ID, we use the notation \textit{"corr"}.

The velocity dispersion in the nuclear regions remains high after including the secondary component (upper panels), with the values for \nii~and \ha~velocity dispersion reduced by $\sim$20--30$\%$ and $\sim$30--40$\%$, respectively, from the values obtained from single Gaussian fits which might indicate the velocity dispersion is intrinsically high. The persisting high velocity dispersion might also suggest that our approach is simplistic and that using more than two Gaussian components possibly can account for the excess better and can reduce the estimated velocity dispersion more. However, increasing the number of components introduces more free parameters, and we do not have enough constraints or resolution to fit more than two components. Moreover, we notice that the largest reduction in velocity dispersion (middle panels) occurs in spaxels where adding the secondary component significantly changes the velocity of the main lines (lower panels).

Therefore, based on the bottom panels of Fig. \ref{fig:vel-sigma-R12R26} we defined the following conditions to create a mask region where single Gaussian fits are not reliable, and values from the run with the secondary component will be used:
\begin{enumerate}[label=\roman*., align=left, leftmargin=*]
    \item $V_{\rm [N \ II]\lambda6583}-V_{\rm [N \ II]\lambda6583,corr}<-4 $ \kms 
    \item $V_{\rm H\alpha}-V_{\rm H\alpha,corr}>5$ \kms
    \item $-6\arcsec\leq \Delta\alpha \leq1.2\arcsec$ and $-3\arcsec\leq \Delta\delta \leq4\arcsec$
\end{enumerate}
The application of the conditions above is shown in the left panel of Fig. \ref{fig:mask_kinematics_excess}. The spatial cuts are applied to discard the isolated spaxels outside the region of the elevated velocity dispersion in the nucleus where the velocity correction is low. The velocity limits are chosen to maximise the area where the secondary component can be reliably fitted. By replacing within the defined mask region,  the measurements from the single Gaussian fits with those from fits with the secondary component included, we created a combined data set of corrected flux, velocity and velocity dispersion. The resultant maps are presented in Fig. \ref{fig:HA_NII_gaskinematics_corrected}. For both \ha~and \nii, the apparent change in flux, velocity and velocity dispersion maps are not significant. However, the accurate corrected values are crucial for our further analysis in Sect. \ref{sec:method}. Hereafter, we use measurements from the combined data set. \oiii~can still be represented by the maps in Fig. \ref{fig:HA_NII_OIII_gaskinematics} since single Gaussian fits were not affected by the presence of the excess.

We present the flux (F) and kinematic maps of the secondary component (indexed by [N\,II]w) within the mask region in the right panel of Fig. \ref{fig:mask_kinematics_excess}, both absolute and relative to the main \nii~component. Note that few spaxels within the mask show inconsistent values compared to the main gradient of flux and kinematic measurements. This implies that the secondary component fits in those spaxels are not reliable and those spaxels are discarded in the analysis below. The secondary component has high flux at the centre of the galaxy that extends towards the northeast side of the mask, coinciding with the high velocity dispersion of the \ha~and \nii~components. In addition, the flux ratio of the secondary component to the main component increases away from the centre. In the velocity map of the secondary component, we see a gradient consistent with rotation, but after we subtract the disk rotation, the secondary component appears blueshifted by 300--400\kms~with respect to the main component, and the blueshift decreases away from the galaxy centre. The velocity dispersions of the secondary and main components are similar, though there is a slight indication that in spaxels where the contribution from the secondary component is the strongest, its velocity dispersion is smaller than that of the main component by up to 20\%.

\noindent
\section{Analysis of the derived kinematics and fluxes}
\label{sec:method}
\subsection{Searching for coherent kinematic structures in deviations from circular motion}
\label{sec:methods-vcirc}

Coherent structures in velocity maps can be best traced when the rotational component of motion is removed from the data. It is most commonly done by subtracting the line-of-sight (LOS) velocity of a simple rotating disk from the observed LOS velocity. The result is displayed in a spatial map, which is called the \textit{residual velocity map}. Coherent structures in the residual velocity map will indicate large-scale motions of gas that depart from circular trajectories. In particular, shocks in gas result in velocity discontinuities, and large-scale shocks are expected to appear as coherent structures in the residual velocity map.

\begin{figure}
    \centering
    \includegraphics[width=0.9\columnwidth]{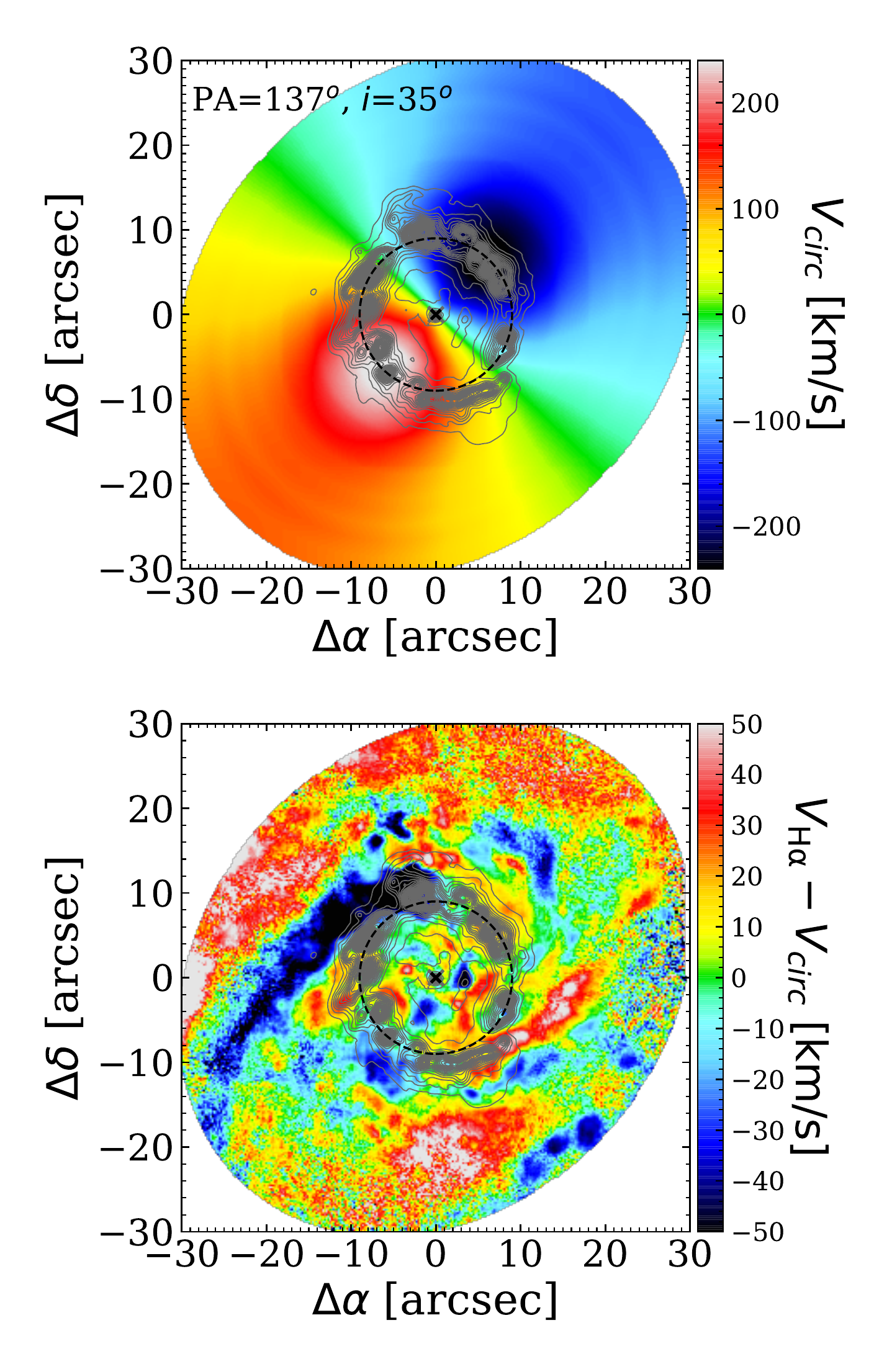}
    \caption{\textit{top:} LOS velocity of the rotating disk fitted to observed \ha~velocity field with PA of the LON $=137{\degr}$ and $i=35{\degr}$. \textit{bottom:} Residual velocities after subtracting the LOS velocity of the modelled disk from the observed \ha~velocity. Grey contours highlight the \ha~emission. The nuclear ring is represented by the black dashed circle. The centre of the galaxy is marked with a cross.}
    \label{fig:Vha-Vcirc}
\end{figure}

In Figure \ref{fig:Vha-Vcirc}, we demonstrate what we define as coherent structures in a residual velocity map. In its upper panel, we present the LOS velocity of a representative thin flat rotating disk which we constructed using \texttt{Kinemetry} software \citep{Krajnovic_06}, described in detail later in this section. The disk orientation parameters, position angle (PA) of the line of nodes (LON) of 137${\degr}$ and inclination (\textit{i}) of 35${\degr}$, are taken from the literature \citep{Fathi_06}. In the bottom panel of Fig. \ref{fig:Vha-Vcirc}, we present the constructed residual velocity map after subtracting the LOS velocity of the modelled disk from the observed \ha~velocity. The residual velocity map reveals two straight continuous structures, one with positive and one with negative residual velocities, which are roughly parallel to the LON and are offset from the LON to the top left and bottom right. Such large residual velocities appearing over continuous stretches in the velocity map are associated with bar-induced shocks in gas which cause velocity jumps. The shocked gas loses angular momentum and funnels from the outer regions of the galaxy towards the inner regions. As shocks compress the gas, which is well mixed with dust, they also coincide with the loci of the dust lanes. Therefore the alignment of the kinematic structures with large residual velocities shown in Fig. \ref{fig:Vha-Vcirc} and the dust lanes shown in Fig. \ref{fig:1097_vimos_cflux} signifies the presence of large-scale shocks that extend from the outer regions of the bar towards the central regions. Although gas inflow in bar-induced shocks is efficient, in Fig. \ref{fig:Vha-Vcirc} it can be seen that the straight shocks weaken in the central regions. As the result, inflow slows down, and the density of the gas increases which can trigger SF and form stellar substructures like the nuclear rings \citep{Shlosman_1989, Athanassoula_1992, Martini_03, Kim_2012}. This effect is seen in the central $\sim$8\arcsec~of NGC\,1097, where the stretches of large residual velocities are connecting to the nuclear ring (shown in Fig. \ref{fig:Vha-Vcirc}).

Thus, although we have a good understanding of the bar driven gas inflows to the central regions, we lack an understanding of the inflow to the innermost nuclear regions. In this work, we aim to identify the inflow of gas from the nuclear ring towards the nucleus of the galaxy by searching for coherent structures in the residual velocity fields. Since subtracting the LOS velocity of an incorrectly fitted disk from the observed velocity will obscure the real features \citep{van_der_Kruit-Allen-1978}, we need to construct the most accurate disk model. To do this, we must cautiously determine the galaxy centre and disk orientation parameters: PA of the LON  and \textit{i}. Below, we explain in detail the procedures of constructing a flat disk in circular motion.

\begin{figure*}
    \centering
    \includegraphics[width=2\columnwidth]{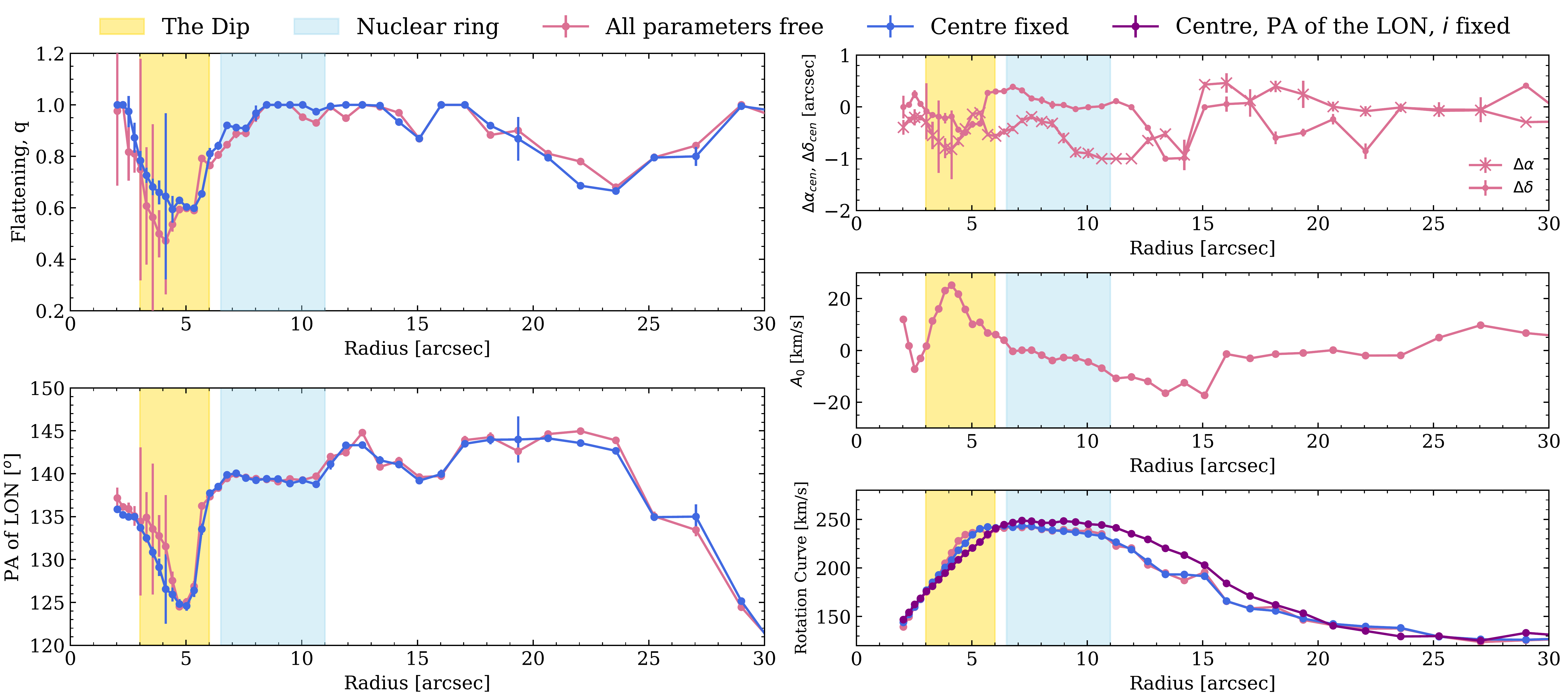}
    \caption{Radial distributions of parameters fitted to the \ha~velocity field with \texttt{Kinemetry}. The blue-shaded region represents the nuclear ring and the yellow-shaded region marks the radial range where PA of the LON and \textit{i} vary significantly (the Dip). Results from the run with all parameters in the \texttt{Kinemetry} fit allowed to vary freely are shown in pink, results for the centre and $A_0$ fixed are shown in blue, and results for the PA of the LON, \textit{i}, centre and $A_0$ fixed are shown in purple. The left panels display the flattening, \textit{q} (top) and the PA of the LON (bottom). The right panels show the coordinates of the kinematic centre (top), the $A_0$ coefficient (middle) and the rotation curve (bottom).}
    \label{fig:Kinemetry_parameters_XCYCincluded}
\end{figure*}
For a flat thin disk in circular motion, the distribution of LOS velocities can be expressed by:
\begin{equation}
    \label{eq:velocityprofile}
    V(R,\psi)=V_{sys}+V_C(R) \ sin(i) \ cos(\psi)
\end{equation}
where $V_{sys}$ and $V_C$ are the systemic and circular velocity respectively, \textit{i} and $\psi$ are the inclination of the disk and azimuthal angle measured counterclockwise from the LON in the galaxy plane, and $R$ is the radius of a circular ring in the galaxy plane. Inclination is directly related to the flattening parameter (or axial ratio) \textit{q} ($cos(i)=q$) of the disk projected on the sky, which is an ellipse with major axis on the LON and ellipticity $\epsilon= 1-q$.

To determine the orientation of the rotating disk model, we employed \texttt{Kinemetry}, which is software specifically designed for IFU data and is a generalisation of surface photometry to higher-order moments of the LOSVD \citep{Krajnovic_06}. \texttt{Kinemetry} performs harmonic expansion along best-fitting ellipses of the optimal major axis’ PA,  optimal ellipticity and centre using Fourier analysis. When applied to the velocity field, it produces the distribution of velocity along ellipses of semi-major axis \textit{a}, K(\textit{a},$\psi$), described by a finite number of harmonic terms:
\begin{equation}
\label{eq:harmonicterms}
    K(a,\psi)=A_0(a)+\sum_{n=1}^{N} A_n(a)sin(n\psi)+B_n(a)cos(n\psi)
\end{equation}
where $\psi$ is the eccentric anomaly, equivalent to that in Equation \ref{eq:velocityprofile}. To determine the best-fitting ellipse, the appropriate harmonic terms are minimised. Under the assumption of emission from a thin disk structure and fixed centre, \texttt{Kinemetry} reduces to the tilted-rings method, which breaks the disk into individual rings with different \textit{i}, PA of the LON and rotational velocity \citep{Rogstad_1974}.

\begin{figure}
    \centering
    \includegraphics[width=0.9\columnwidth]{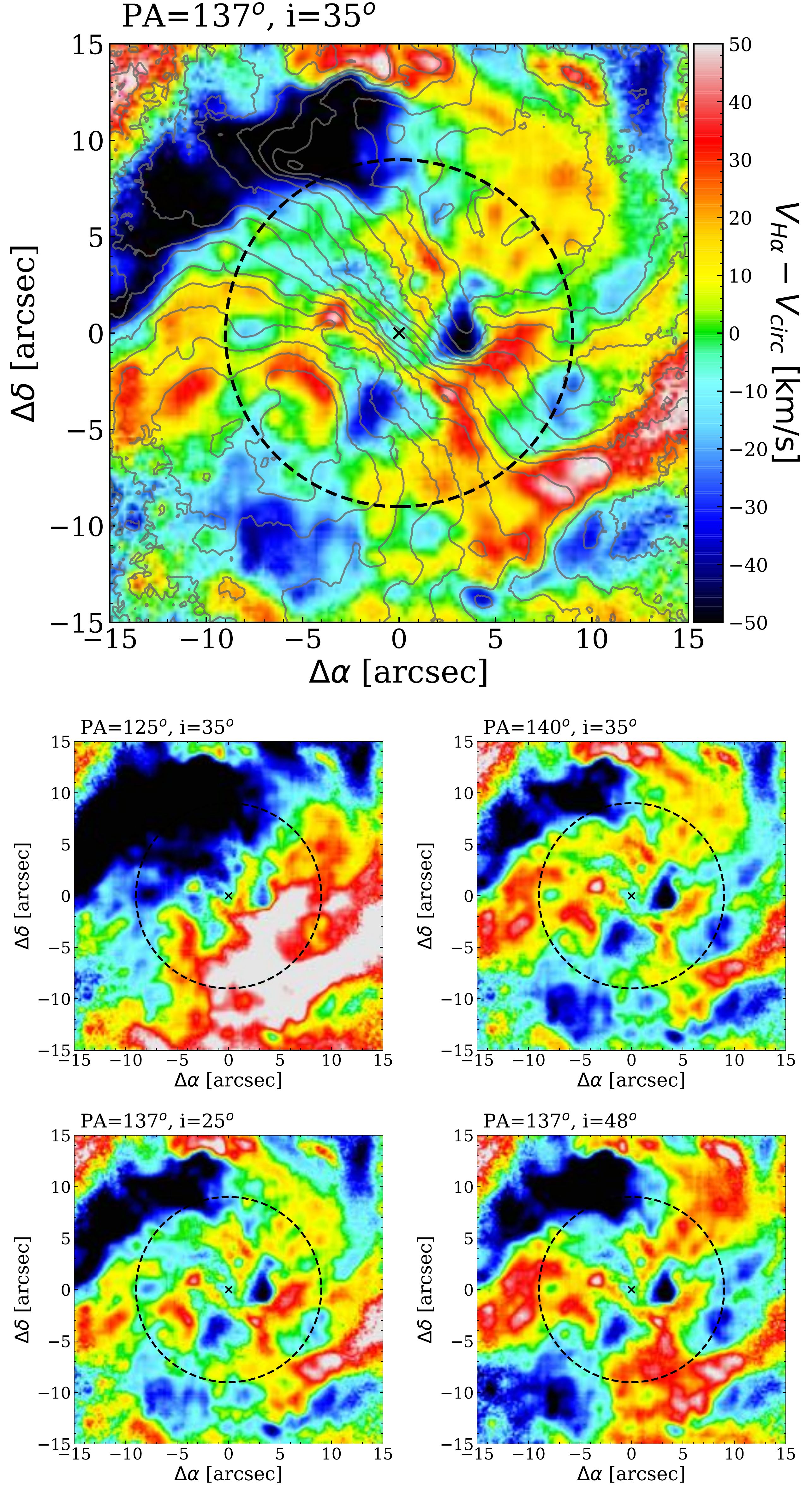}
    \caption{ Residual velocity map of the inner 30\arcsec$\times$30\arcsec. The dashed circle represents the nuclear ring. The centre of the galaxy is marked with a cross. In each panel, we subtract the LOS velocity of the disk modelled with a different combination of PA of the LON and \textit{i}. \textit{top:} PA=137${\degr}$ and \textit{i}=35${\degr}$ ($q=0.82$) as in Fig. \ref{fig:Vha-Vcirc}. The grey contours represent the \ha~velocity field.  \textit{middle:} \textit{i} is fixed at $35{\degr}$ and PA of the LON is varied. \textit{bottom:} PA of the LON is fixed at $137{\degr}$ and \textit{i} is varied.} 
    \label{fig:ranged_q_residuals}
\end{figure}
Since we want to fit the rotating disk to the gas motions, we use the \ha~velocity field as the input for \texttt{Kinemetry}. To ensure the thin flat rotating disk is accurately fitted we performed three \texttt{Kinemetry} fits, starting from one with all orientation parameters allowed to vary and gradually fixing the global orientation parameters in the consecutive fits. In Figure \ref{fig:Kinemetry_parameters_XCYCincluded}, we summarise the estimated parameter distributions from each fit. In determining the global values for the disk orientation parameters, we discard parameters estimated within the inner 2\arcsec~as the estimations within this radius are unreliable because of the insufficient data that ellipses are being fitted to, and because of the impact of the AGN.

We first fitted the velocity field by allowing the orientation parameters of each ellipse, including the centre, to vary freely. As shown in the top right panel of Fig. \ref{fig:Kinemetry_parameters_XCYCincluded}, in the regions inward from the nuclear ring, the horizontal ($\Delta\alpha$) and vertical coordinates ($\Delta\delta$) of the kinematic centre show only small variations ($\sim$0.7\arcsec~for $\Delta\alpha$ and $\sim$0.5\arcsec~for $\Delta\delta$) from the photometric centre. This indicates that in the regions inner to the nuclear ring, the estimated kinematic centre agrees with the photometric centre within $\sim$3 spaxels. Also, as shown in the middle right panel of Fig. \ref{fig:Kinemetry_parameters_XCYCincluded}, the $A_0$ values for ellipses fitted interior to the nuclear ring depart from the systemic velocity of the galaxy by less than 25\kms. Therefore, we reduced the number of free parameters within \texttt{Kinemetry} by fixing the centre of the galaxy to the coordinates of the photometric centre, and the $A_0$ to the velocity at the photometric centre ($-$3.55\kms), and performed a second fit. In this second fit, we intended to estimate a representative PA of the LON and \textit{i} which can accurately describe the ellipses in nuclear regions, and we aimed to build our final circular disk model by fixing the PA of the LON and \textit{i} to these estimated representative values. However, as shown in the left panels of Fig. \ref{fig:Kinemetry_parameters_XCYCincluded}, regardless of keeping the centre free or fixed, in the regions inward from the nuclear ring, the distributions of the PA of the LON and \textit{i} strongly fluctuate within a radial region marked in yellow, which we refer to as \textit{the Dip}. Within the Dip, the PA of the LON from the second fit varies between 125{\degr}--140{\degr}, and \textit{q} varies between 0.65--0.90, which translates to an \textit{i} range of $\sim$25\degr-- 48\degr. While the impact of bar-induced noncircular motions can justify the fluctuation of these parameter values outside the nuclear ring, inward from the nuclear ring one would expect motion close to circular and therefore well fitted by single values for the PA of the LON and \textit{i}. However, this expectation might be simplistic, as NGC\,1097 is proposed to be hosting a nuclear bar \citep{Shaw_1993,Buta_1993} which would impact the gas flow in the regions inward from the nuclear ring.

As we indicate above, in the third \texttt{Kinemetry} fit we want to fix the PA of the LON and \textit{i} to fit a flat disk, but it is challenging to deduce reliable parameter values from such fluctuating distributions. Moreover, in Figure \ref{fig:Vha-Vcirc}, which presents residual velocities for a thin disk fit with parameters taken from the literature \citep[PA of the LON $=137{\degr}$ and $i=35{\degr}$,][]{Fathi_06}, in the region inner to the nuclear ring we notice a periodic azimuthal variation resembling an $m$\,=\,3 mode. Our initial explanation for it was that \textit{i} used in the disk model was higher than the true value, which would give artefacts in the residuals in the form of $m$\,=\,3 harmonic terms \citep{van_der_Kruit-Allen-1978, Franx_Gorkom_Zeeuw_1994, Schoenmakers_1997}. Artefacts can be removed by choosing an \textit{i} value in the disk model that is closer to the actual \textit{i} of the galaxy, but variations in \texttt{Kinemetry} parameters are preventing us from settling on a reliable \textit{i}.

Therefore, to find the \textit{i} value for which the residuals of the \texttt{Kinemetry} fits are free from the $m$\,=\,3 azimuthal variations, we investigated the effect of changing the \textit{i} and PA of the LON on residual velocities by running \texttt{Kinemetry} first with a fixed PA of the LON at 137${\degr}$ and different \textit{q} (0.7, 0.9), and then with a fixed \textit{q} at 0.82 ($i\approx$35\degr) and different PA of the LON (125${\degr}$, 140${\degr}$). We constructed the corresponding residual velocity maps and examined the residuals, as we show in Fig. \ref{fig:ranged_q_residuals}, to deduce which parameter set is best for the final disk model. When the disk is modelled with the PA of the LON $125{\degr}$ and \textit{i} $35{\degr}$, we see the $m$\,=\,1 harmonic terms in the residual velocity map as an effect of wrongly fitted PA of the LON, with one side of the map having mainly positive velocities and the other side having mainly negative velocities \citep{van_der_Kruit-Allen-1978}. However, these artefacts can be largely eliminated by adopting the optimal PA of the LON. We also see $m$\,=\,3 harmonic terms in all panels. They are strongest in the bottom right panel where we notice both positive and negative residual velocity regions extending radially from the centre beyond the nuclear ring. We must emphasise that no matter what inclination is used in the disk model, $m$\,=\,3 azimuthal variation in the residual velocity is only minimised and not eliminated. The \textit{i} within the range of 25${\degr}$-- 35${\degr}$ minimises its amplitude, yet we do not see a clear minimum. 

The range of acceptable parameter values agrees well with the parameters in the literature: the kinematic position angle, PA=135${\degr}$, derived in \citet{Storchi-Bergmann_1996}  and \citet{Emsellem_01}; inclination, \textit{i=34${\degr}$}, derived in \citet{Storchi-Bergmann_03} for the nuclear accretion disk, and \textit{i=35${\degr}$} derived in \citet{Fathi_06} from exponential thin disk model. Therefore we conclude that the disk parameters, PA=137${\degr}$ and $i$=35$\degr$, taken from \citet{Fathi_06} and applied in Fig. \ref{fig:Vha-Vcirc}, are optimal. Although the parameters, PA=122${\degr}$ and $i$=48${{\degr}}$ adopted by \citet{Lang_20} differ from our optimal parameters, we show in Fig. \ref{fig:ranged_q_residuals} that each of these parameters would give artefacts in the residual velocity.

Due to the smoother nature of the stellar velocity field compared to the gas velocity field, it is generally expected to obtain more continuous parameter distributions when fitting a disk to the stellar velocity field. However, the large velocity dispersion inward from the nuclear ring of NGC\,1097 (Fig. \ref{figapp:stellarmaps}) reduces the LOS velocity values over an elliptical region inconsistent with a projected disk in an acceptable inclination range, and therefore is likely to result in an incorrect fit, which we indeed observed when using \texttt{Kinemetry}. Therefore, parameters estimated by fitting the stellar velocity field would not be reliable inward from the nuclear ring. On the other hand, the PA of the LON and \textit{i} that we derive with \texttt{Kinemetry} for the stellar velocity field outside the nuclear ring is consistent with our adopted values, although the PA of the LON diverges at larger radii, likely because of the influence of the bar. Furthermore, disk orientation parameters cannot be reliably obtained from large-scale photometry due to the presence of companion and asymmetric spiral arms.

Since, even for the optimal parameters, a flat disk in circular motion does not provide a good fit to the data, because periodic azimuthal variations of residual velocities persist in those regions, we revisited the LOS velocity fields of each emission line group to check if there are any signs of these features directly in the observed velocity fields. In the top panel of Fig. \ref{fig:ranged_q_residuals}, we superimpose \ha~velocity contours on the residual velocity map to show that in the regions inward from the nuclear ring, the \ha~velocity field shows wiggles at loci coinciding with two most prominent negative residual velocity regions that we interpreted above as part of the $m$\,=\,3 harmonic terms. This indicates that the features which we initially interpreted as artefacts of the fit arise from real features present in the data. Hence, in Sect. \ref{sec:BPT}, we focus on examining the physical processes in the regions inner to the nuclear ring, which may give rise to some of the kinematic structures observed in the velocity map.

\begin{figure*}
    \centering
    \includegraphics[width=2\columnwidth]{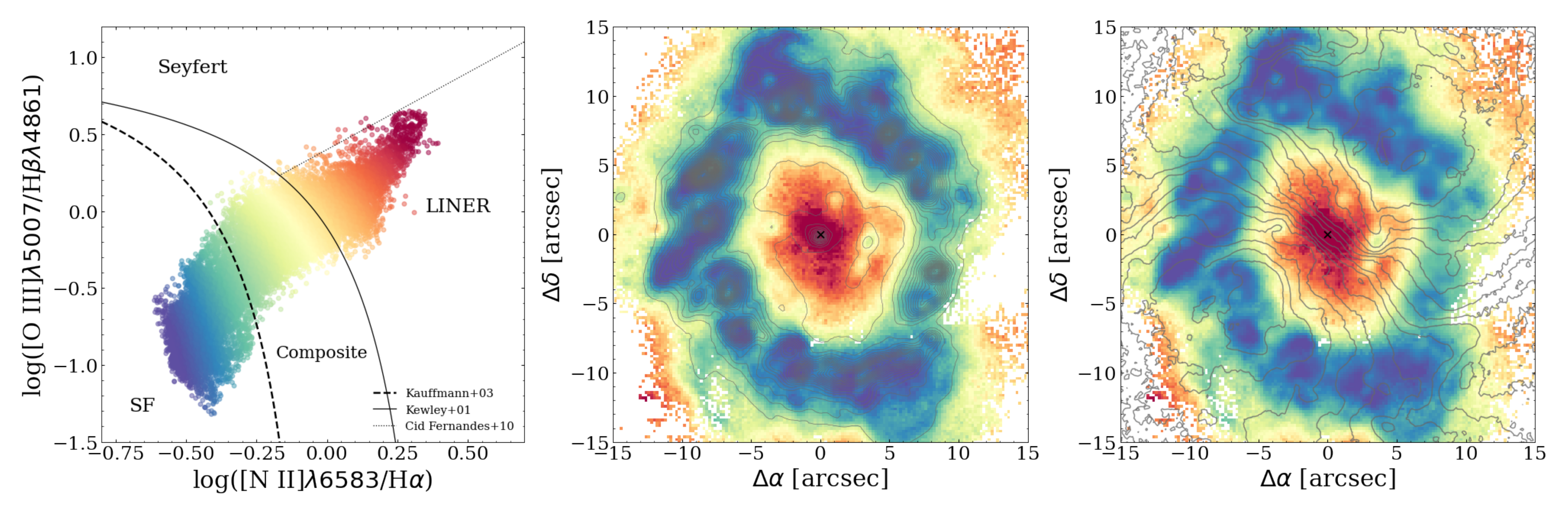}
    \caption{ \textit{left:} The BPT diagram of the inner 30\arcsec$\times$30\arcsec~region. The SNR$=$10 and AON$=$7 thresholds are applied. Each point represents one spaxel. Dashed, solid and dotted lines represent the \citet{Kauffmann_03}, \citet{Kewley_06} and \citet{Cid_Fernandes_10} fits respectively which establish the dominating ionisation mechanisms. The colour coding is based on the distance from the \citet{Kewley_06} line. \textit{middle \& right:} Maps of the inner 30\arcsec$\times$30\arcsec, with colour coding of the spaxels taken from the left panel. Contours represent \ha~flux (middle) and LOS \ha~velocity (right). }
    \label{fig:BPT_spatial}
\end{figure*}

\subsection{Dominating ionisation mechanisms in the centre of NGC\,1097}
\label{sec:BPT}
To determine the dominating source of ionisation from optical emission lines, such as AGN or star-forming regions, one can use the Baldwin, Phillips \& Terlevich (BPT) diagram \citep{Baldwin_1981}. In the left panel of Fig. \ref{fig:BPT_spatial}, we show the BPT diagram for the inner 30\arcsec$\times$30\arcsec~region of NGC\,1097. Each point on the diagram corresponds to a spaxel in the data, with its position determined by the fluxes measured for that spaxel. We mask spaxels with poor flux measurements by applying an amplitude-over-noise (AON) ratio \citep{Sarzi_06} threshold of 7, which excludes spaxels with AON$\leq$7 from the BPT diagram, choosing a smaller threshold would give noisy results already occurring outside the nuclear ring in Fig. \ref{fig:BPT_spatial}. This allows us to focus on the emission that is coming from the nuclear ring and the region inner to it. To distinguish AGN from star-forming regions, we used the \citealp{Kewley_01a} (hereafter Kewley+01) relation which indicates what line ratios can be produced by stellar photoionization, as points above the Kewley+01 line likely originate from the AGN activity or shocks. However, the Kewley+01 relation does not take into account the composite emission which has a contribution from SF, hot old low-mass stars and AGN. Therefore, to include the composite emission, we used the modified Kewley+01 line from \citealp{Kauffmann_03} (hereafter Kauffmann+03) so that the region between Kaufmann+03 and Kewley+01 lines is the location of the composite sources, and points below the Kaufmann+03 line indicate emission purely coming from SF. We also separated the AGN region into two classes, LINER and Seyfert region, based on the relation given by \citet{Cid_Fernandes_10}. 

Each point in the BPT diagram represents a spaxel and is colour coded according to its distance from the Kewley+01 line. This colour coding is then used in the spatial map to highlight which sources of ionisation dominate in what regions (Fig. \ref{fig:BPT_spatial} central and right panels). We also included \ha~flux and velocity contours on the spatial maps to combine the BPT diagram results with kinematics and light distribution.

In the nuclear ring, the emission is strongly dominated by SF, while the regions inward from the nuclear ring predominantly show LINER emission. However, three locations in the regions inward from the nuclear ring ($\sim$5\arcsec~east and west and $\sim$2\arcsec~northwest from the centre) show composite emission due to contribution from SF and have high \ha~flux (see also Fig. \ref{fig:HA_NII_OIII_gaskinematics}). The region of composite emission to the west has the highest contribution from SF (is closest to the Kaufmann+03 line in the BPT diagram) and it appears to be part of a structure with elevated SF that extends from the nuclear ring directly north of it towards a smaller region of the composite emission to the south. The \ha~velocity field throughout the extent of this structure is strongly disturbed (Fig. \ref{fig:BPT_spatial} right panel), which can be caused by stellar outflows associated with elevated SF. Moreover, this structure also coincides with the strong negative velocity residual (Fig. \ref{fig:ranged_q_residuals}), suggesting that the negative velocity residuals, which we initially interpreted as artefacts caused by adopting the wrong inclination of the fitted disk, arise from real features present in the data. The region of composite emission to the east shows a weaker SF component than the region to the west. This region is also associated with a disturbed \ha~velocity field, and unlike the region to the west which is extended tangentially in the north-south direction, it is curved towards the nucleus of the galaxy. Lastly, the region of the composite emission to the northwest has the weakest contribution from SF, and the \ha~velocity field in that region is only mildly disturbed. Compared to the other two regions, this region appears to be more isolated yet it still clearly shows in the velocity residuals in Fig. \ref{fig:ranged_q_residuals}.

\subsection{Searching for coherent kinematic differences between distinct emission line groups}
\label{sec:method-vres_emissionlines}

\begin{figure*}
    \includegraphics[width=0.33\textwidth]{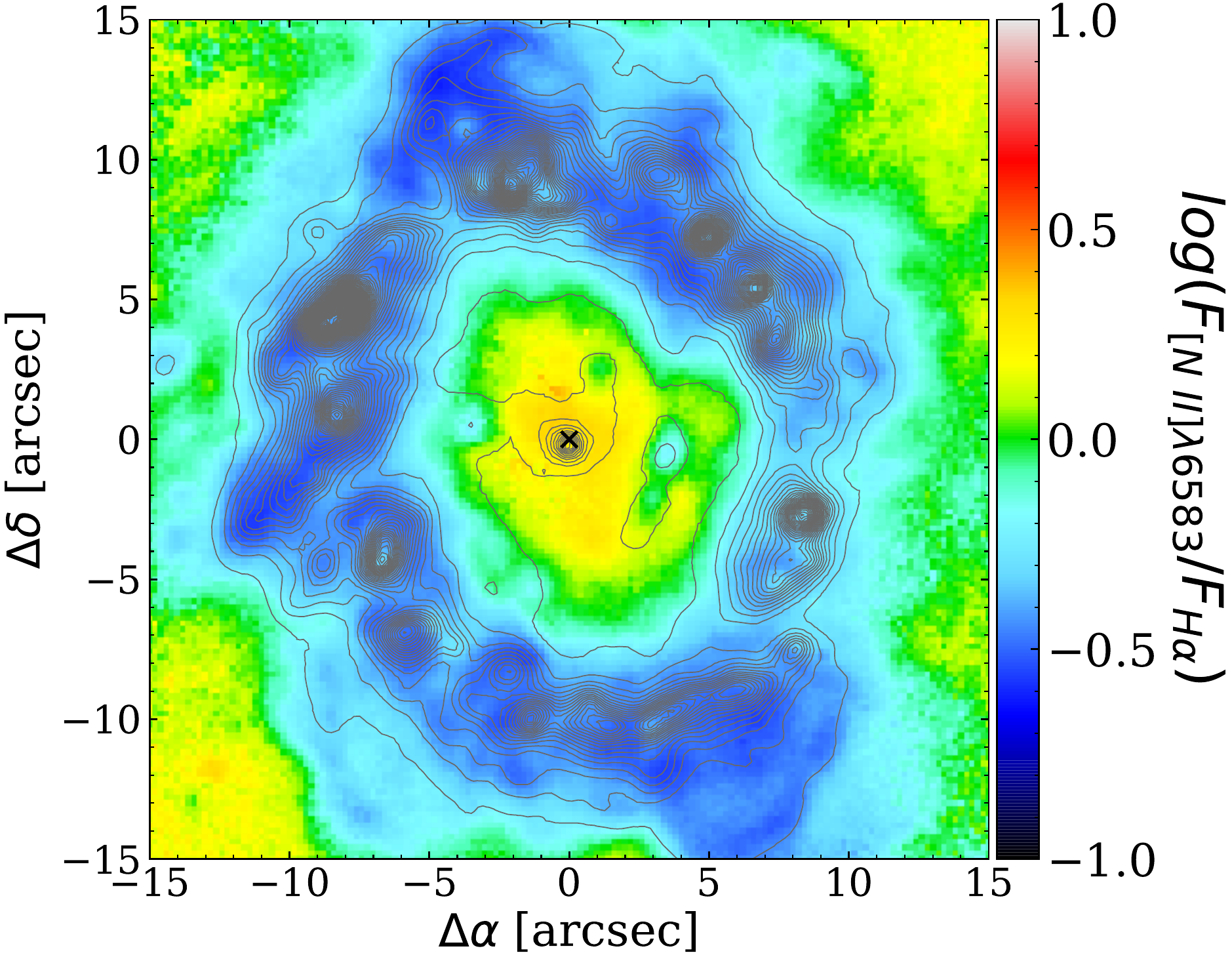}\hfill
    \includegraphics[width=0.33\textwidth]{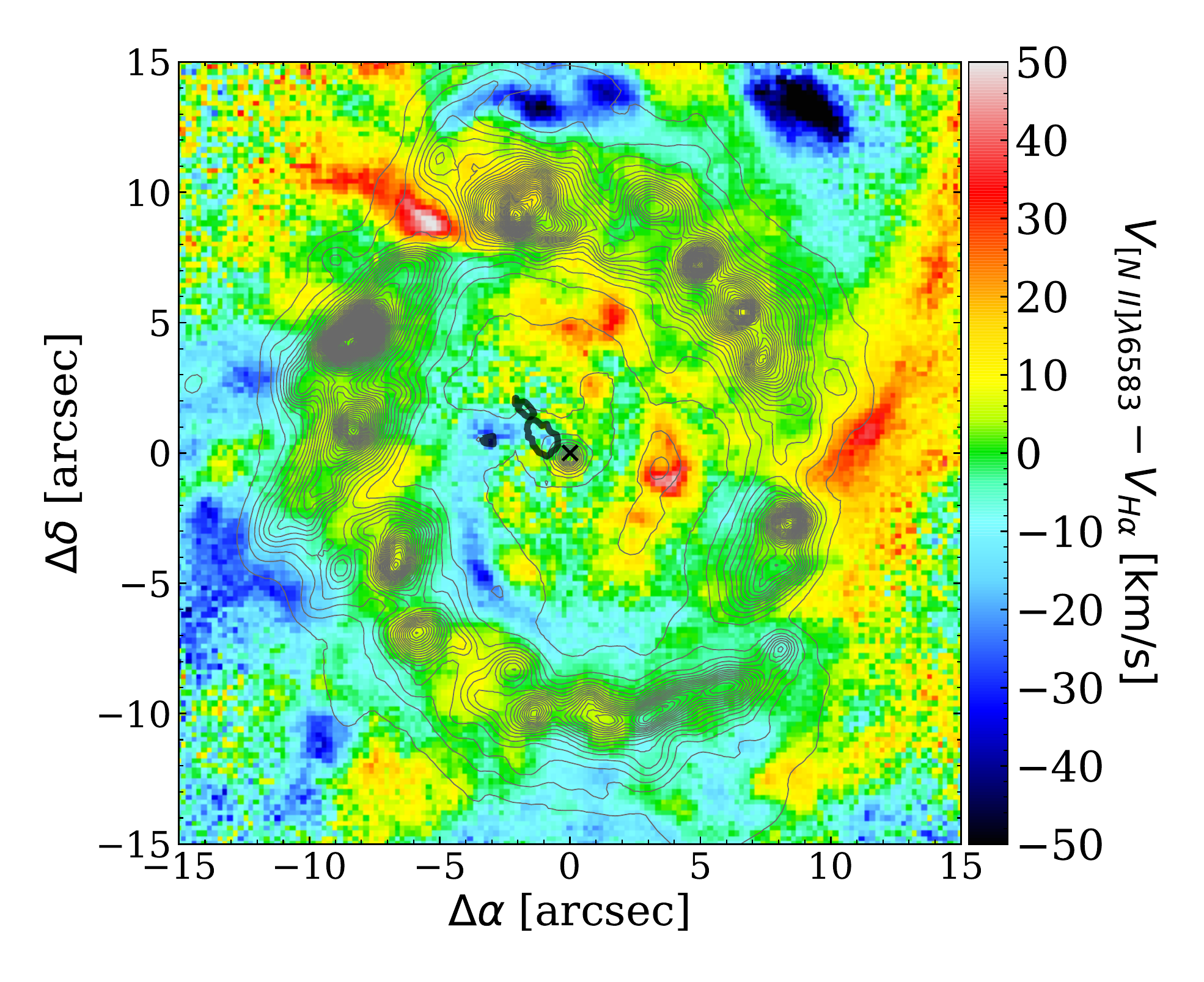}\hfill
    \includegraphics[width=0.33\textwidth]{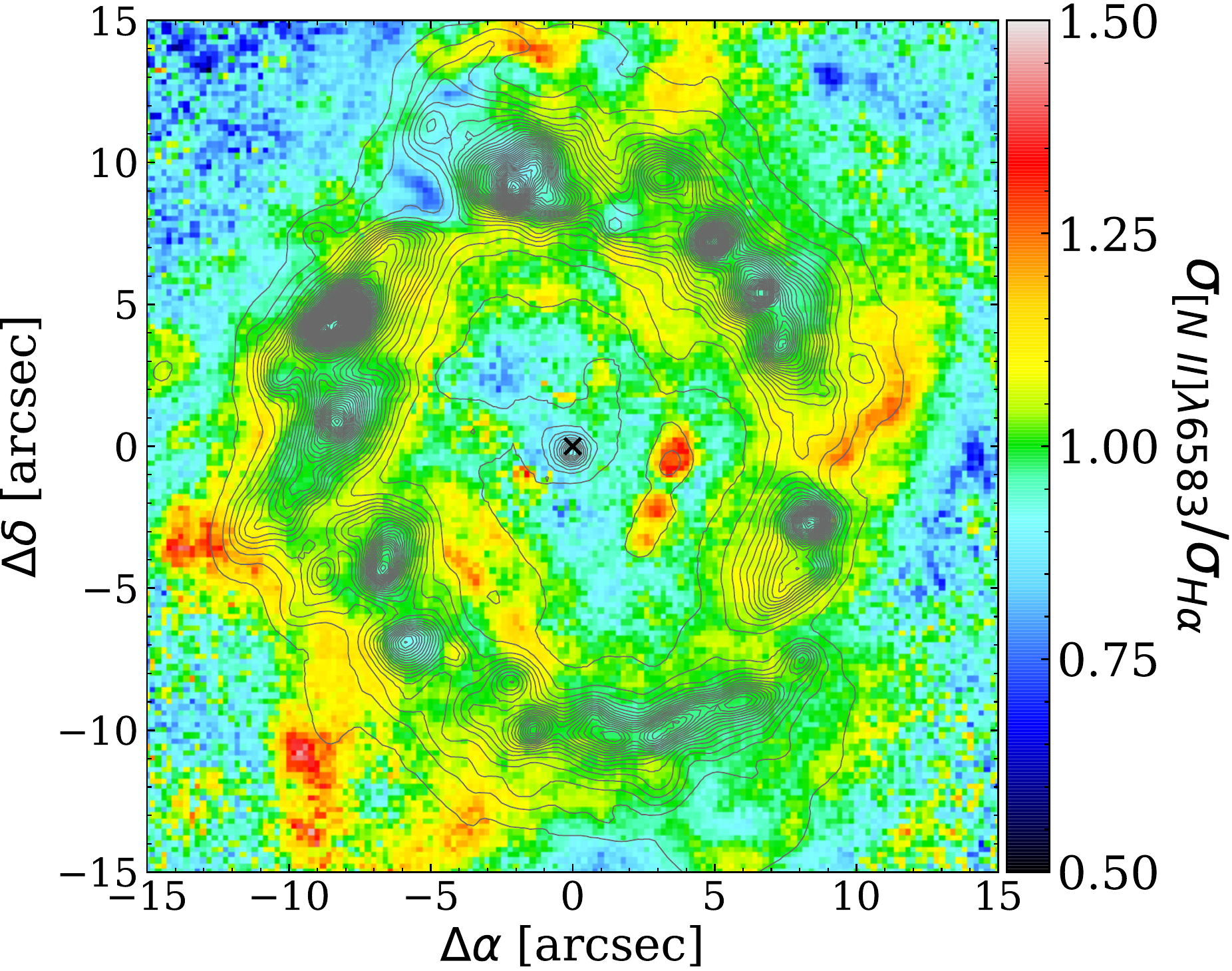}
    \caption{ \textit{left:} Line flux ratio of \nii~and \ha. \textit{middle:} The velocity difference map, which is constructed by subtracting the LOS velocity of \ha~from the LOS velocity of \nii. Black contour marks the region where the velocity difference between \nii~and \ha~was more than 30\kms for single Gaussian fits. \textit{right:} Velocity dispersion ratio of \nii~and \ha. In all panels, grey contours highlight the \ha~emission.}
    \label{fig:Vres_emissionlines}
\end{figure*}

We showed in Sect. \ref{sec:methods-vcirc} that velocity residuals after subtracting a flat disk in circular motion do not provide conclusive information about coherent kinematic structures inner to the nuclear ring. Here, we explore what kinematic information can be extracted from the data without relying on kinematic or mass models. In Fig. \ref{fig:Vres_emissionlines}, we compare the LOSVD moments of the Balmer and low ionisation lines by constructing the maps of [N\,II]$\lambda$6583/\ha~line flux ratio (left panel), velocity difference (middle panel) and velocity dispersion ratio (right panel). 

\nii~emission, with higher ionisation potential, may predominantly originate from post-shock gas, while \ha~emission, mostly generated by stellar photoionization, is less affected by the shock. Therefore, in the regions where shocks are present, the observed [N\,II]$\lambda$6583/\ha~line flux ratios should be elevated. As shown in the left panel of Fig. \ref{fig:Vres_emissionlines}, [N\,II]$\lambda$6583/\ha~is the lowest in the nuclear ring where SF dominates the emission, and it is the highest in the nucleus where the AGN dominates. There are no other regions inward from the nuclear ring with elevated \nii~emission. Instead, we notice regions of elevated \ha~flux associated with enhanced SF (see Sect. \ref{sec:BPT}), which might be obscuring the possible enhancement in \nii~in the same region if SF is triggered by shocks. Nevertheless, if \nii~around shocks originates mostly from the post-shock gas, large-scale shocks can manifest themselves by velocity differences between \nii~and \ha~that are coherent over large distances, as explained below. The velocity of each tracer evaluated in a spaxel is the first moment of the LOSVD arising from integrating the emission along the LOS and over the spaxel area. If this includes regions of gas with different physical conditions which move with respect to each other, the integrated LOSVD recorded in a spaxel should vary between the tracers, as relative contributions to line emission from regions with different physical conditions differ between emission lines. Consequently, the value of its first moment may differ for each tracer. 

A good example of two neighbouring gas regions that differ in physical conditions and move with respect to one another are regions upstream and downstream from a shock in gas. Commonly temperature and pressure are higher in the post-shock gas than in gas upstream from the shock, and jump conditions in the shock impose a velocity difference between these two regions. Hence, for a spaxel which records emission from pre-shock and post-shock regions, the derived \nii~and \ha~velocities will differ, as the former will be more representative of the post-shock region. Then, if large-scale shocks are present, they will appear as continuous structures in the velocity difference map, in which the velocity difference takes consistently positive (or negative) values. In the velocity difference map shown in the middle panel of Fig. \ref{fig:Vres_emissionlines}, we notice  a coherent blue feature of negative velocity difference spiralling out from the nucleus towards the southeast side of the nuclear ring, hereafter called the \textit{blue spiral arm}. If it is a signature of shock in gas, it may be a spiral shock \citep{Maciejewski_2004} extending inwards from the large-scale straight shocks in the bar of NGC\,1097 \citep{Fathi_06}. The largest velocity difference in the blue spiral occurs at the location where the blue spiral curves towards the nucleus and is cospatial with the \ha~enhancement to the east and with enhanced SF (see Fig. \ref{fig:BPT_spatial}). This may indicate that the enhanced SF is associated with the inner part of the spiral shock.

Furthermore, shock conditions give rise to random motions, which increase the velocity dispersion of \nii~in the post-shock regions. This can possibly be seen in the southeast region of the blue spiral arm, where the velocity dispersion of \nii~is higher than \ha~but in the region where the blue spiral curves towards the nucleus, the velocity dispersion of \nii~is only slightly above $\sim$1. In the region $\sim$5\arcsec west from the nucleus we notice the largest velocity dispersion ratio, which is cospatial with the largest velocity differences to the west, and the elevated [N\,II]$\lambda$6583/\ha~line flux ratios in the same region, as well as the enhanced SF (Sect. \ref{sec:BPT}), but this feature is less spatially coherent than the blue spiral arm.

Although the observed velocity differences between emission lines could result from obscuration, \ha~and \nii~emission wavelengths differ only by 20\AA, so differential extinction is negligible. While incorrect line fitting can also cause the difference in the derived velocities between tracers, we minimise that possibility by addressing the non-Gaussianity of the emission lines by including the secondary component of \nii~line  in spaxels where the excess emission was present, as we explain in Sect. \ref{sec:excess}. In fact, fitting emission lines with single Gaussian components resulted in the largest velocity difference between \ha~and \nii~in spaxels where the excess in the spectra was present. In the middle panel of Figure \ref{fig:Vres_emissionlines}, we mark with a black contour the region where the velocity difference ranged from $-$30 to $-$50 \kms for single Gaussian fits. Including the secondary component reduced this difference to $\sim$10\kms. The spectra throughout the blue spiral arm, which is outside the contour region, are well-fitted with single Gaussians, which indicates that the velocity difference between emission lines in those regions is not caused by the incorrect fitting.

\section{Discussion}
\label{sec:discussion}

\subsection{Elliptical flow pattern in regions inward from the nuclear ring}
\label{sec:discussion-non-circ}
\begin{figure*}
    \centering
    \includegraphics[width=1.7\columnwidth]{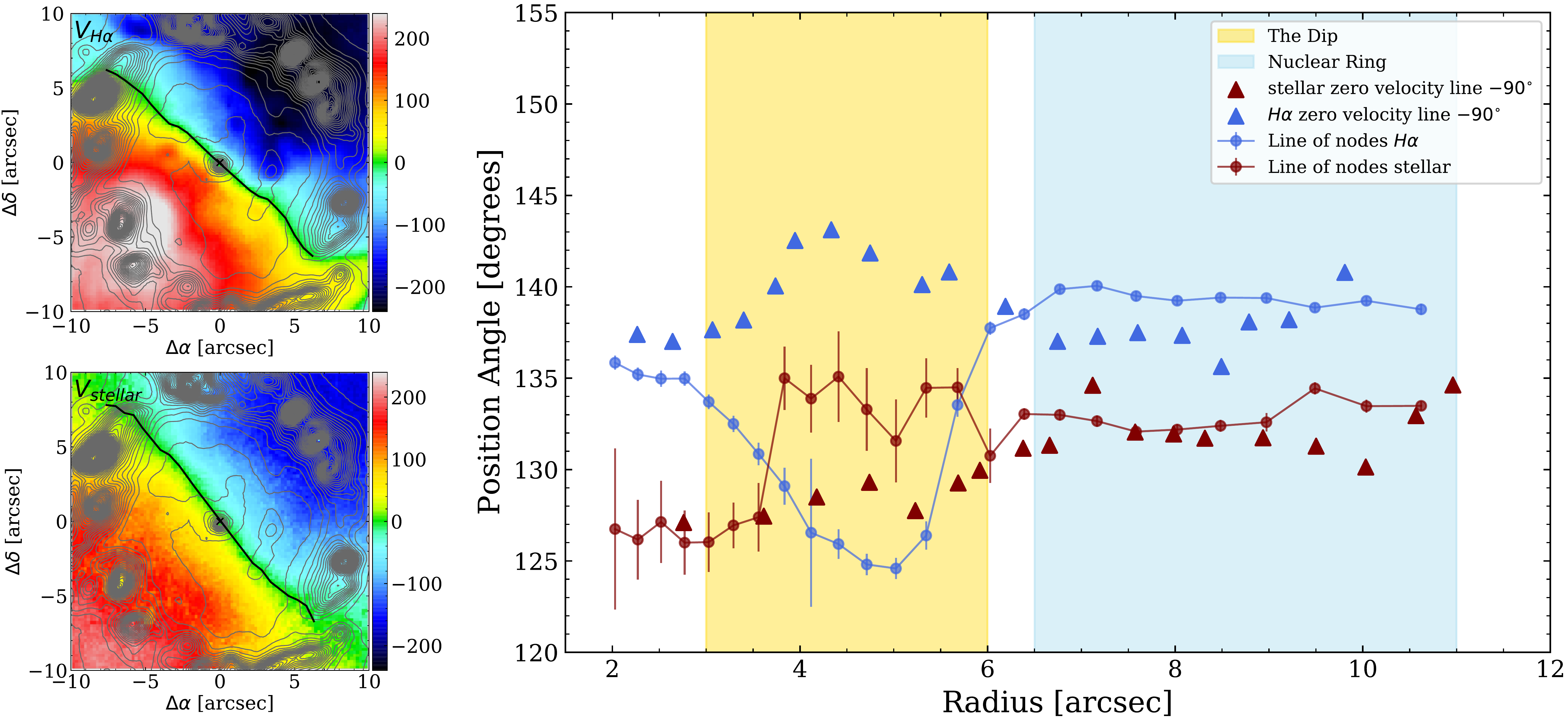}
    \caption{\textit{left:} \ha~(top) and stellar (bottom) LOS velocity maps. The black line represents the corresponding zero velocity line fit. Grey contours represent the \ha~emission. The centre of the galaxy is marked with a cross. \textit{right:} Distributions of the PA of the LON obtained from \texttt{Kinemetry} and PA of the zero velocity line. For the stars the distribution of the PA of the LON is plotted in small red circles and the PA of the zero velocity line in red triangles. For gas, the PA of the LON is plotted in small blue circles and the PA of the zero velocity line in blue triangles. The blue-shaded region represents the nuclear ring, and the yellow-shaded region represents the Dip defined in Sect. \ref{sec:methods-vcirc}.}
    \label{fig:PA_combined}
\end{figure*}

We based our studies of residual velocities (Sect. \ref{sec:methods-vcirc}) on the assumption that inward from the nuclear ring the flow of unperturbed gas is circular. However, we showed that in this region \texttt{Kinemetry} cannot accurately constrain the parameters of the unperturbed circular motion in a thin flat disk. The possible reason for this is that the unperturbed flow is either not flat or not circular there. In order to obtain meaningful residuals, one needs to first correctly identify the underlying flow.

If the flow is intrinsically circular, the PA of the LON and PA of the zero velocity line are perpendicular, but this is not the case if the global flow is intrinsically not circular (e.g. elliptical; see the appendix of \citealp{Mazzalay_14}). Here, we test whether in the regions inward from the nuclear ring the underlying flow is intrinsically noncircular, by determining the angle between the LON and the zero velocity line, where the distribution of the PA of the LON is inferred from \texttt{Kinemetry}.

In order to derive the distribution of the PA of the zero velocity line,
for each column of spaxels in the velocity field (corresponding to constant Right Ascension), we identified the spaxel with the velocity closest to the systemic velocity of the galaxy. We then connected these spaxels to produce the fit of the zero velocity line. Further, we determined the PA of each spaxel on the zero velocity line which gives us the distribution of the PA of the zero velocity line with radius. We applied these steps on both stellar and \ha~velocity fields. We expect that the angle between the LON and the zero velocity line will be less affected by non-circular motion for stars than for gas because high velocity dispersion in stars hides the underlying non-circularity.

In the left panels of Fig. \ref{fig:PA_combined}, we show the \ha~and stellar LOS velocity maps and the corresponding zero velocity line fitted on each map. In the right panel, we present the individual distributions of the PA of the LON and the PA of the zero velocity line for both velocity fields. In the figure, we show the PA of the zero velocity line values after subtracting 90{\degr}, which should be equal to the PA of the LON for circular motion. The distributions of the PA of the LON and PA of the zero velocity line for stars roughly align with each other over the whole radial range, differing slightly inside the Dip (see Sect. \ref{sec:methods-vcirc}) by at most $\sim$5{\degr}. The values of the PA of the LON and PA of the zero velocity line for \ha~velocity field roughly converge in the nuclear ring, but within the Dip, the two differ by $\sim$15{\degr}--20{\degr}. This implies that in the regions inward from the nuclear ring, the underlying orbits of the gas flow are intrinsically noncircular, which prevents \texttt{Kinemetry} from accurately constraining the disk orientation. 

Thus, assuming an intrinsically circular flow of gas may compromise the search for coherent kinematic structures in the velocity residuals, and an oval or elliptical flow can be a better representation of the gas flow. Such assumption of an oval flow is allowed by the \texttt{DiskFit} software \citep{Spekkens_Sellwood_07}, which is specifically developed for bar-like and oval distortions and allows fitting non-axisymmetric flow patterns to two-dimensional velocity fields. Although exploring this approach is plausible, it is outside the scope of this work.

\subsection{Evidence from the MUSE data for nuclear spiral shocks}
\label{sec:discussion-NACO-NB}

\begin{figure}
    \includegraphics[width=0.9\columnwidth]{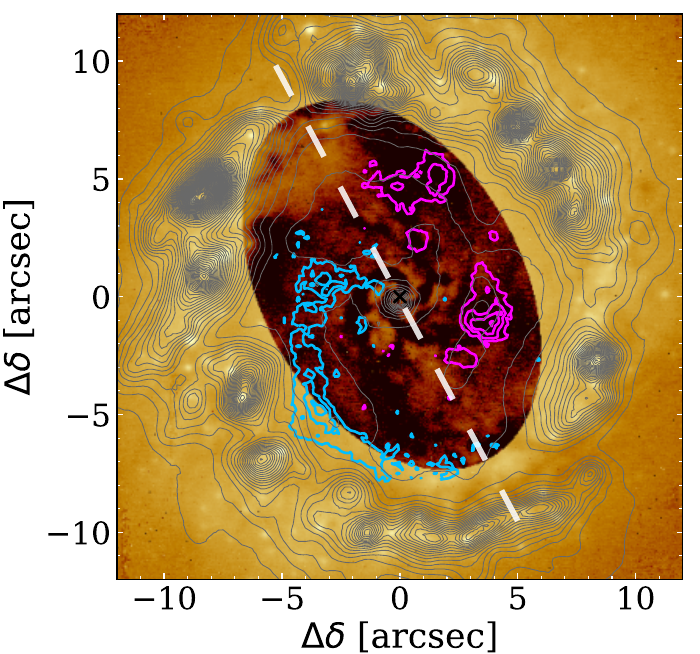}
    \caption{NACO J-band image from \citet{Prieto_05} with the light profile of a simple ellipse model subtracted inside the oval. The white dashed line represents the proposed nuclear bar with PA=28${\degr}$ and length\,$\approx$\,20\arcsec~\citep{Quillen_95}. Blue and pink contours represent the blue and red spiral arms in the velocity difference map (Fig. \ref{fig:Vres_emissionlines}). Grey contours represent the \ha~emission.}
    \label{fig:NACO}
\end{figure}

Studying the near-infrared images from the adaptive optics assisted NACO camera/spectrograph of the VLT, \citet{Prieto_05} found dust extinction in the form of a complex three-arm filamentary structure in the regions inner to the nuclear ring. The spiral arms of this structure are reaching to the centre from the north, south and west. Moreover, in good agreement with \citet{Prieto_05}, a similar three-arm network is observed in the images from the Advanced Camera for Surveys (ACS) of the HST \citep{Fathi_06}. Hereafter we call this structure the \textit{nuclear dust filaments}. In Figure \ref{fig:NACO}, we show the NACO image of \citet{Prieto_05}. Inside the oval area, the image displays the residuals after subtracting the light profile of a simple ellipse model. In the figure, we include a dashed line to represent the proposed nuclear bar of the galaxy \citep{Shaw_1993,Buta_1993}, with a major axis PA of 28\degr \citep{Quillen_95}. We notice that the northern and southern arms of the nuclear dust filaments are roughly parallel to this nuclear bar, which is consistent with the expected geometry of dust lanes in a bar when the dust lanes are associated with gas inflow in shocks. The arm to the west of the nucleus is not readily associated with the nuclear bar.

In Figure \ref{fig:NACO}, in blue contours, we outline the blue spiral arm extending from the inner southeast part of the nuclear ring towards the nucleus which is revealed in the velocity difference map (Fig. \ref{fig:Vres_emissionlines}, middle panel). The region of the blue spiral where it curves towards the nucleus is cospatial with the region of the composite emission directly to the east of the nucleus, which is associated with enhanced SF, and with the disturbed \ha~velocity field (Fig. \ref{fig:BPT_spatial}). If the velocity difference is caused by the shock, as we argued in Sect. \ref{sec:method-vres_emissionlines}, then the blue spiral arm can be a nuclear spiral shock in which gas inflow to the nucleus occurs. For bisymmetric mass distribution, such shock should have a symmetric red component on the other side of the nucleus \citep{Canzian_1993,Schoenmakers_1997} in the velocity difference map. In Fig. \ref{fig:Vres_emissionlines}, middle panel, we notice a red structure at $\sim$5\arcsec~west of the nucleus extending from the nuclear ring towards the south, outlined in pink in Fig. \ref{fig:NACO}. This structure is not curved towards the nucleus and does not have as defined spiral continuity as the blue spiral arm, but it has the highest velocity differences, it is cospatial with the region of the composite emission to the west, which has the highest contribution from SF, and it coincides with the strongest distortions of the \ha~velocity field. Hereafter, we call this structure the \textit{red spiral arm}. We hypothesise that in the red spiral arm the gas density is higher compared to the region to the east, which leads to higher SF there, whose consequences, particularly outflows, dominate the shock signature there, while the SF is less dominant in the blue spiral. 

\begin{figure}
    \centering
    \includegraphics[width=1\columnwidth]{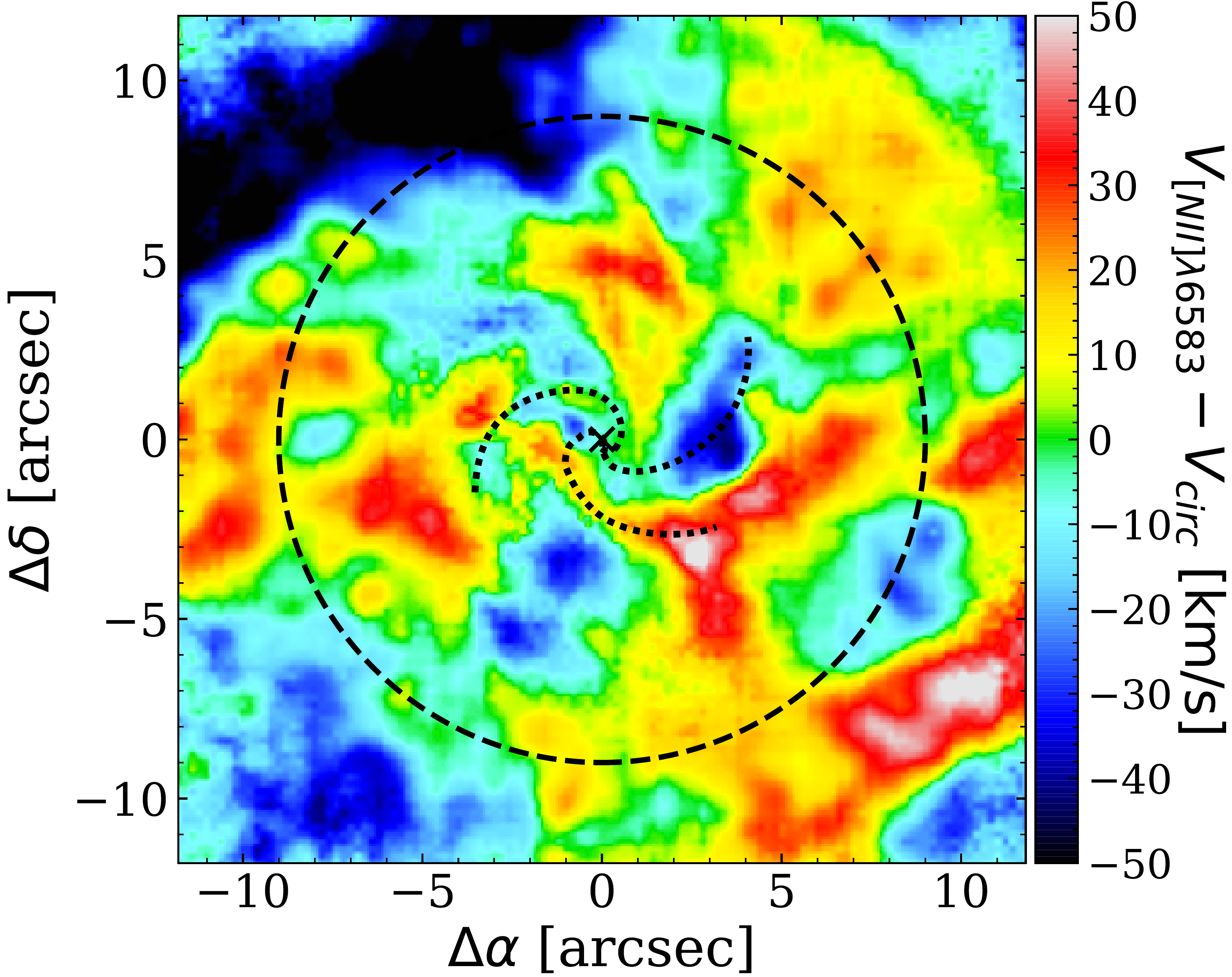}
    \caption{Residual velocities after subtracting the LOS velocity of the disk model from the observed \nii~velocity field. The black dotted lines represent the kinematic spiral network of \citet{Fathi_06}. The black dashed circle represents the nuclear ring. The centre of the galaxy is marked with a cross.}
    \label{fig:Fathi_residualmap}
\end{figure}

The blue and red spiral arms are shifted $\sim$2-3\arcsec~ downstream from the northern and southern arms of the nuclear dust filaments. Hence, the combination of the structures observed in kinematics and morphology might be the imprint of the gas flowing to the nucleus under forcing from the nuclear bar. However, it is unclear whether the kinematic arms are directly associated with the dust filaments.

Hydrodynamical and kinematic modelling \citep{Davies_09,Fathi_13} support a net molecular gas inflow to the nucleus, channelled via the arms of the nuclear dust filaments of \citet{Prieto_05}. Using data from the Gemini Multi-Object Spectrograph (GMOS) on Gemini South Telescope, \citet{Fathi_06} studied the \nii~kinematics within the inner 15\arcsec$\times$7\arcsec~region of NGC\,1097, and revealed a  three-arm spiral network in residual velocities centred on the galaxy nucleus with velocity amplitudes reaching $\sim$20$\%$ of the rotational velocity, consistent with the nuclear spiral models of \citet{Maciejewski_2004}. Moreover, \citet{Davies_09} studied the H$_2$ kinematics within the inner 4\arcsec$\times$4\arcsec~region and could resolve a two-arm kinematic spiral structure in their residual velocities. Benefiting from the large FoV of MUSE, we can trace these observed kinematic structures in our residual velocity map beyond their spatial coverage. We do not find any correlation between the structures found in the residual velocities by \citet{Davies_09} and our results, but the FoV of SINFONI is much smaller than the region to which we fitted the disk -- Note that the SINFONI data have higher spatial resolution than the MUSE data, which might possibly make it harder to find a match as well.

Since \citet{Fathi_06} used \nii~data in their analysis, to compare our findings, we fitted our disk model with fixed PA of the LON=137${\degr}$ and \textit{i}=35${\degr}$ (see Sect. \ref{sec:methods-vcirc}) to \nii~velocity field. The resulting residual velocity map is presented in Fig. \ref{fig:Fathi_residualmap}, which also includes the proposed three-arm kinematic spiral of \citet{Fathi_06}. We do not find a coherent spiral network extending from the galaxy nucleus towards outer radii. Arguably, the proposed arm of the kinematic spiral to the south appears connected with a coherent structure in our residual velocity map, which might be extending from the nuclear ring towards the nuclear regions, but instead of curving towards the nucleus, it continues to the northeast. As already seen in \citet{Fathi_06} this arm runs along the arm of the nuclear dust filaments to the west. Hence, although in the region inward from the nuclear ring a three-arm spiral network is present in morphology \citep{Prieto_05,Fathi_06}, we do not find evidence of a spiral network in the residual velocities of the ionised gas. As we explained in Sects. \ref{sec:methods-vcirc}, \ref{sec:BPT} and \ref{sec:discussion-non-circ}, the data inward from the nuclear ring do not allow an accurate fit of a flat disk in circular motion, therefore residual velocities have to be interpreted with caution. On the other hand, in this work we find indications of nuclear spiral shocks in velocity difference maps.

\subsection{Multi-phase AGN outflows in the nuclear regions}
\label{sec:discussion-outflows}

As confirmed by a number of studies, the gas funnelling into the nucleus of NGC\,1097 feeds its LINER core \citep{Davies_09,vandeVen_Fathi_2010, Fathi_13, Izumi_13, Martin_15,Izumi_17,Prieto_19}. Some of those studies also attempt to compute the mass inflow rate ($\dot{M}$) from the nuclear ring of the galaxy into its central black hole. For example, \citet{vandeVen_Fathi_2010} estimated the ionised $\dot{M}$ at $\sim$70pc from the galaxy nucleus as $\sim$0.011$M_{\odot}yr^{-1}$, \citet{Fathi_13} estimated a net $\dot{M}$ as $\sim$0.2$M_{\odot}yr^{-1}$ at $\sim$40pc and $\sim$0.6$M_{\odot}yr^{-1}$ at $\sim$100pc from the galaxy nucleus by using combined molecular and ionised gas observations, and \citet{Davies_09} estimated a net $\dot{M}$ of 0.06$M_{\odot}yr^{-1}$ to the central few tens of parsecs regions. Magnetic fields in the region inner to the nuclear ring that can affect gas inflow rates have been observed \citep{Lopez-Rodriguez_21,Hu_22}.

It is known that the inflow of gas into the central black hole can feed the AGN and trigger powerful outflows \citep{Garcia-Burillo-05,Davies_14}. Even though this phenomenon is more commonly seen in Seyfert nuclei \citep{Garcia_Burillo_2014,Alonso-Herrero_19,Audibert-Combes-2020}, outflows can still be powered in less luminous cores \citep{Nemmen_06,Santoro_20}, as supported by recent multi-wavelength studies of nearby LINER galaxies \citep{Marquez_17, Hermosa_Munoz_22}. In this section, we discuss the imprint of ionised gas outflows in the nucleus of NGC\,1097, which may be triggered due to the inflow of gas into the central black hole.

As we discussed in Sect. \ref{sec:excess}, we noticed a high velocity dispersion region extending from the centre of the galaxy towards $\sim$3\arcsec~northeast (see Fig. \ref{fig:HA_NII_OIII_gaskinematics}). One reason for the high velocity dispersion values is the excess to the blue wing of \nii~line, which is affecting the line profiles and the Gaussian fits (see Fig. \ref{fig:representative_excess_spectra}). However, even after modelling the excess as a separate secondary emission component, the velocity dispersion of the main component remained high (see Fig. \ref{fig:vel-sigma-R12R26}). Contribution from the secondary component to the total \nii~flux appears to increase with radius and its velocity is high and blueshifted compared to the main \nii~line (Fig. \ref{fig:mask_kinematics_excess}).

X-ray/Chandra observations adopted by \citet{Prieto_19}, highlight the X-ray-rich nucleus of the galaxy along with the energetic extension blasting from the nucleus towards the northeast, northwest and southwest. The northeast X-ray extent coincides with the high velocity dispersion region. A similar correlation between ionised \ha~emission and soft X-ray observations has been reported by previous studies on nearby LINER galaxies \citep{Marquez_17, Hermosa_Munoz_22}. Moreover, high velocity dispersion of various molecular gas tracers is also observed in the regions inward from the nuclear ring \citep{Izumi_17,Leroy_2021b}. 

We hypothesise that the strong disturbances in the kinematics of the multiple gas phases and ionised gas disturbances coinciding with the loci of X-ray emission imply the presence of AGN outflows, triggered by the gas inflows into the black hole. In particular, emission from the secondary component could be coming from the jet-like outflow, whose interaction with the disk increases the velocity dispersion in the latter.

\section{Summary \& Conclusions} 
\label{sec:conclude}
In this pilot study, we presented methods that can be used in the search for gas inflow in centres of galaxies, and we applied them to a detailed study of the kinematics of ionised gas in the inner 60\arcsec$\times$60\arcsec~region of NGC\,1097. Using data from MUSE/VLT, we derived kinematic maps for emission lines representing three distinct groups: Balmer, low ionisation and high ionisation lines.

In the regions inward from the nuclear ring, we improved the quality of line fitting by including a secondary component to the \nii~emission line in order to account for its asymmetric line profile. The morphology and kinematics of gas giving rise to the secondary component, when compared with previous multi-wavelength studies, indicate that this emission could be related to the outflows from the AGN.

Shocks in gas, especially those of coherent span over large distances, like straight shocks in bars and spiral shocks, have long been associated with inflow, therefore we argue that gas inflow can be revealed by coherent kinematic structures that large-scale shocks impress in kinematic maps. These structures are best highlighted when the revolving motion around the centre is removed, but we showed that subtracting the circular velocity of a flat disk may not reveal coherent kinematic structures when gas is not moving on circular orbits and its revolution around the centre is better approximated by motion along an oval. In addition, local perturbations, like outflows from SF regions can distort the appearance of coherent kinematic structures induced by large-scale shocks in gas.

Instead of subtracting the circular velocity of a modelled disk, the contribution from the revolving motion around the centre can be minimised in a model-independent way by subtracting velocity maps of two different emission lines, which should highlight differences in kinematics between two tracers.
 
The map of velocity difference between \nii~and \ha~constructed for the central kiloparsec of NGC\,1097 reveals a coherent structure in the form of a spiral, extending from the inner southeast part of the nuclear ring to the north and curving towards the nucleus (the "blue spiral arm"), whose southeastern region shows higher velocity dispersion of \nii~compared to \ha. We associate the coherent velocity difference in this structure with shock conditions and argue for the blue spiral arm being a large-scale shock in the gas. In the velocity difference map, the blue spiral arm appears to have a symmetric companion on the other side of the nucleus (the "red spiral arm"). Although less coherent than the blue spiral arm, it coincides with the region of high SF and strongly disturbed \ha~velocity field, which indicates that if the red spiral arm is also a shock in gas, it is strongly affected by SF activity. The bisymmetric geometry of the blue and red spiral arms is consistent with predictions from nuclear spiral models. The kinematic arms that we found in NGC\,1097 run along the southern and northern spiral arms of the nuclear dust filaments, located $\sim$3\arcsec~downstream from them. Together, they appear aligned well with the nuclear bar, which suggests that the nuclear bar can play a role in gas inflow in the centre of NGC\,1097.

The methodology developed in this pilot study provides new means of finding coherent kinematic structures in gas associated with shocks and inflows; by applying it to the diverse galaxy sample of nearby galaxies contained in CBS, in the following paper we intend to determine how common gas inflows are in coherent kinematic structures.

\section*{Acknowledgements}
TK acknowledges the joint studentship support from Liverpool John Moores University the Faculty of Engineering and Technology, and the Science Technology Facilities Council. TK acknowledges the studentship support from European Southern Observatory. TK thanks Dr Almudena Prieto for sharing the processed NACO J-band image of NGC\,1097, Professor Philip A. James for useful comments on the manuscript, Professor Eric Emsellem for suggestions and discussions on data analysis, Dr Francesco Belfiore for valuable help on general enquiries with the Data Analysis Pipeline, and Dr Adrian Bittner for help on the GIST pipeline. TK especially thanks Alonso Luna Ruiz Fernández for his continuous support and encouragement.
\newline
JN acknowledges funding from the European Research Council (ERC) under the European Union’s Horizon 2020 research and innovation programme (grant agreement No. 694343).

\section{Data Availability}
MUSE data underlying this work is accessed via the data archive of the European Southern Observatory, under proposal ID:0097.B-0640 (PI: Dimitri Gadotti) provided by the TIMER collaboration, at \url{http://archive.eso.org/wdb/wdb/adp/phase3$\_$main/form}.
\label{lastpage}

\bibliographystyle{mnras}
\bibliography{references} 

\begin{thebibliography}{}
\makeatletter
\relax
\def\mn@urlcharsother{\let\do\@makeother \do\$\do\&\do\#\do\^\do\_\do\%\do\~}
\def\mn@doi{\begingroup\mn@urlcharsother \@ifnextchar [ {\mn@doi@}
  {\mn@doi@[]}}
\def\mn@doi@[#1]#2{\def\@tempa{#1}\ifx\@tempa\@empty \href
  {http://dx.doi.org/#2} {doi:#2}\else \href {http://dx.doi.org/#2} {#1}\fi
  \endgroup}
\def\mn@eprint#1#2{\mn@eprint@#1:#2::\@nil}
\def\mn@eprint@arXiv#1{\href {http://arxiv.org/abs/#1} {{\tt arXiv:#1}}}
\def\mn@eprint@dblp#1{\href {http://dblp.uni-trier.de/rec/bibtex/#1.xml}
  {dblp:#1}}
\def\mn@eprint@#1:#2:#3:#4\@nil{\def\@tempa {#1}\def\@tempb {#2}\def\@tempc
  {#3}\ifx \@tempc \@empty \let \@tempc \@tempb \let \@tempb \@tempa \fi \ifx
  \@tempb \@empty \def\@tempb {arXiv}\fi \@ifundefined
  {mn@eprint@\@tempb}{\@tempb:\@tempc}{\expandafter \expandafter \csname
  mn@eprint@\@tempb\endcsname \expandafter{\@tempc}}}

\bibitem[\protect\citeauthoryear{{Allison}, {Sadler}  \& {Meekin}}{{Allison}
  et~al.}{2014}]{Allison_14}
{Allison} J.~R.,  {Sadler} E.~M.,   {Meekin} A.~M.,  2014, \mn@doi [\mnras]
  {10.1093/mnras/stu289}, \href
  {https://ui.adsabs.harvard.edu/abs/2014MNRAS.440..696A} {440, 696}

\bibitem[\protect\citeauthoryear{{Alonso-Herrero} et~al.,}{{Alonso-Herrero}
  et~al.}{2019}]{Alonso-Herrero_19}
{Alonso-Herrero} A.,  et~al., 2019, \mn@doi [\aap]
  {10.1051/0004-6361/201935431}, \href
  {https://ui.adsabs.harvard.edu/abs/2019A&A...628A..65A} {628, A65}

\bibitem[\protect\citeauthoryear{{Athanassoula}}{{Athanassoula}}{1992}]{Athanassoula_1992}
{Athanassoula} E.,  1992, \mn@doi [\mnras] {10.1093/mnras/259.2.345}, \href
  {https://ui.adsabs.harvard.edu/abs/1992MNRAS.259..345A} {259, 345}

\bibitem[\protect\citeauthoryear{{Audibert} et~al.,}{{Audibert}
  et~al.}{2019}]{Audibert_19}
{Audibert} A.,  et~al., 2019, \mn@doi [\aap] {10.1051/0004-6361/201935845},
  \href {https://ui.adsabs.harvard.edu/abs/2019A&A...632A..33A} {632, A33}

\bibitem[\protect\citeauthoryear{{Audibert} et~al.,}{{Audibert}
  et~al.}{2020}]{Audibert-Combes-2020}
{Audibert} A.,  et~al., 2020, arXiv e-prints, \href
  {https://ui.adsabs.harvard.edu/abs/2020arXiv201109133A} {p. arXiv:2011.09133}

\bibitem[\protect\citeauthoryear{{Audibert}, {Combes}, {Garc{\'\i}a-Burillo}
  \& {Dasyra}}{{Audibert} et~al.}{2021}]{Audibert_21}
{Audibert} A.,  {Combes} F.,  {Garc{\'\i}a-Burillo} S.,   {Dasyra} K.,  2021,
  \mn@doi [IAU Symposium] {10.1017/S1743921320002239}, \href
  {https://ui.adsabs.harvard.edu/abs/2021IAUS..359..307A} {359, 307}

\bibitem[\protect\citeauthoryear{{Bacon} et~al.,}{{Bacon}
  et~al.}{2010}]{Bacon_10}
{Bacon} R.,  et~al., 2010, in {McLean} I.~S.,  {Ramsay} S.~K.,   {Takami} H.,
  eds,  Society of Photo-Optical Instrumentation Engineers (SPIE) Conference
  Series Vol. 7735, Ground-based and Airborne Instrumentation for Astronomy
  III. p. 773508, \mn@doi{10.1117/12.856027}

\bibitem[\protect\citeauthoryear{{Baldwin}, {Phillips}  \&
  {Terlevich}}{{Baldwin} et~al.}{1981}]{Baldwin_1981}
{Baldwin} J.~A.,  {Phillips} M.~M.,   {Terlevich} R.,  1981, \mn@doi [\pasp]
  {10.1086/130766}, \href
  {https://ui.adsabs.harvard.edu/abs/1981PASP...93....5B} {93, 5}

\bibitem[\protect\citeauthoryear{{Bittner} et~al.,}{{Bittner}
  et~al.}{2019}]{Bitner_19}
{Bittner} A.,  et~al., 2019, \mn@doi [\aap] {10.1051/0004-6361/201935829},
  \href {https://ui.adsabs.harvard.edu/abs/2019A&A...628A.117B} {628, A117}

\bibitem[\protect\citeauthoryear{{Bittner} et~al.,}{{Bittner}
  et~al.}{2020}]{Bittner_20}
{Bittner} A.,  et~al., 2020, \mn@doi [\aap] {10.1051/0004-6361/202038450},
  \href {https://ui.adsabs.harvard.edu/abs/2020A&A...643A..65B} {643, A65}

\bibitem[\protect\citeauthoryear{{Bosma}}{{Bosma}}{1978}]{Bosma_1978}
{Bosma} A.,  1978, PhD thesis, University of Groningen, Netherlands

\bibitem[\protect\citeauthoryear{{Bowen}, {Chelouche}, {Jenkins}, {Tripp},
  {Pettini}, {York}  \& {Frye}}{{Bowen} et~al.}{2016}]{Bowen_16}
{Bowen} D.~V.,  {Chelouche} D.,  {Jenkins} E.~B.,  {Tripp} T.~M.,  {Pettini}
  M.,  {York} D.~G.,   {Frye} B.~L.,  2016, \mn@doi [\apj]
  {10.3847/0004-637X/826/1/50}, \href
  {https://ui.adsabs.harvard.edu/abs/2016ApJ...826...50B} {826, 50}

\bibitem[\protect\citeauthoryear{{Buta} \& {Crocker}}{{Buta} \&
  {Crocker}}{1993}]{Buta_1993}
{Buta} R.,  {Crocker} D.~A.,  1993, \mn@doi [\aj] {10.1086/116695}, \href
  {https://ui.adsabs.harvard.edu/abs/1993AJ....106..939B} {106, 939}

\bibitem[\protect\citeauthoryear{{Canzian}}{{Canzian}}{1993}]{Canzian_1993}
{Canzian} B.,  1993, \mn@doi [\apj] {10.1086/173095}, \href
  {https://ui.adsabs.harvard.edu/abs/1993ApJ...414..487C} {414, 487}

\bibitem[\protect\citeauthoryear{{Cappellari}}{{Cappellari}}{2017}]{Cappellari_2017}
{Cappellari} M.,  2017, \mn@doi [\mnras] {10.1093/mnras/stw3020}, \href
  {https://ui.adsabs.harvard.edu/abs/2017MNRAS.466..798C} {466, 798}

\bibitem[\protect\citeauthoryear{{Cappellari} \& {Copin}}{{Cappellari} \&
  {Copin}}{2003}]{Cappellari_Copin_03}
{Cappellari} M.,  {Copin} Y.,  2003, \mn@doi [\mnras]
  {10.1046/j.1365-8711.2003.06541.x}, \href
  {https://ui.adsabs.harvard.edu/abs/2003MNRAS.342..345C} {342, 345}

\bibitem[\protect\citeauthoryear{{Cappellari} \& {Emsellem}}{{Cappellari} \&
  {Emsellem}}{2004}]{Cappellari-emsellem-2004}
{Cappellari} M.,  {Emsellem} E.,  2004, \mn@doi [\pasp] {10.1086/381875}, \href
  {https://ui.adsabs.harvard.edu/abs/2004PASP..116..138C} {116, 138}

\bibitem[\protect\citeauthoryear{{Cid Fernandes}, {Stasi{\'n}ska}, {Mateus}  \&
  {Vale Asari}}{{Cid Fernandes} et~al.}{2011}]{Cid_Fernandes_10}
{Cid Fernandes} R.,  {Stasi{\'n}ska} G.,  {Mateus} A.,   {Vale Asari} N.,
  2011, \mn@doi [\mnras] {10.1111/j.1365-2966.2011.18244.x}, \href
  {https://ui.adsabs.harvard.edu/abs/2011MNRAS.413.1687C} {413, 1687}

\bibitem[\protect\citeauthoryear{{Davies}, {Maciejewski}, {Hicks}, {Tacconi},
  {Genzel}  \& {Engel}}{{Davies} et~al.}{2009}]{Davies_09}
{Davies} R.~I.,  {Maciejewski} W.,  {Hicks} E.~K.~S.,  {Tacconi} L.~J.,
  {Genzel} R.,   {Engel} H.,  2009, \mn@doi [\apj]
  {10.1088/0004-637X/702/1/114}, \href
  {https://ui.adsabs.harvard.edu/abs/2009ApJ...702..114D} {702, 114}

\bibitem[\protect\citeauthoryear{{Davies} et~al.,}{{Davies}
  et~al.}{2014}]{Davies_14}
{Davies} R.~I.,  et~al., 2014, \mn@doi [\apj] {10.1088/0004-637X/792/2/101},
  \href {https://ui.adsabs.harvard.edu/abs/2014ApJ...792..101D} {792, 101}

\bibitem[\protect\citeauthoryear{{Diehl} \& {Statler}}{{Diehl} \&
  {Statler}}{2006}]{Diehl_06}
{Diehl} S.,  {Statler} T.~S.,  2006, \mn@doi [\mnras]
  {10.1111/j.1365-2966.2006.10125.x}, \href
  {https://ui.adsabs.harvard.edu/abs/2006MNRAS.368..497D} {368, 497}

\bibitem[\protect\citeauthoryear{{Donnari} et~al.,}{{Donnari}
  et~al.}{2021}]{Donnari_21}
{Donnari} M.,  et~al., 2021, \mn@doi [\mnras] {10.1093/mnras/staa3006}, \href
  {https://ui.adsabs.harvard.edu/abs/2021MNRAS.500.4004D} {500, 4004}

\bibitem[\protect\citeauthoryear{{ESO Press Release}}{{ESO Press
  Release}}{2004}]{PR_release_NGC1097}
{ESO Press Release} 2004, {The President and the Galaxy}, ESO Press Release,
  12/2004

\bibitem[\protect\citeauthoryear{{Ellison}, {Nair}, {Patton}, {Scudder},
  {Mendel}  \& {Simard}}{{Ellison} et~al.}{2011}]{Ellison_11}
{Ellison} S.~L.,  {Nair} P.,  {Patton} D.~R.,  {Scudder} J.~M.,  {Mendel}
  J.~T.,   {Simard} L.,  2011, \mn@doi [\mnras]
  {10.1111/j.1365-2966.2011.19195.x}, \href
  {https://ui.adsabs.harvard.edu/abs/2011MNRAS.416.2182E} {416, 2182}

\bibitem[\protect\citeauthoryear{{Emsellem}, {Greusard}, {Combes}, {Friedli},
  {Leon}, {P{\'e}contal}  \& {Wozniak}}{{Emsellem} et~al.}{2001}]{Emsellem_01}
{Emsellem} E.,  {Greusard} D.,  {Combes} F.,  {Friedli} D.,  {Leon} S.,
  {P{\'e}contal} E.,   {Wozniak} H.,  2001, \mn@doi [\aap]
  {10.1051/0004-6361:20000523}, \href
  {https://ui.adsabs.harvard.edu/abs/2001A&A...368...52E} {368, 52}

\bibitem[\protect\citeauthoryear{{Emsellem}, {Goudfrooij}  \&
  {Ferruit}}{{Emsellem} et~al.}{2003}]{Emsellem_03}
{Emsellem} E.,  {Goudfrooij} P.,   {Ferruit} P.,  2003, \mn@doi [\mnras]
  {10.1046/j.1365-2966.2003.07050.x}, \href
  {https://ui.adsabs.harvard.edu/abs/2003MNRAS.345.1297E} {345, 1297}

\bibitem[\protect\citeauthoryear{{Emsellem} et~al.,}{{Emsellem}
  et~al.}{2022}]{Emsellem_22}
{Emsellem} E.,  et~al., 2022, \mn@doi [\aap] {10.1051/0004-6361/202141727},
  \href {https://ui.adsabs.harvard.edu/abs/2022A&A...659A.191E} {659, A191}

\bibitem[\protect\citeauthoryear{{Erwin} et~al.,}{{Erwin}
  et~al.}{2021}]{Erwin_21}
{Erwin} P.,  et~al., 2021, \mn@doi [\mnras] {10.1093/mnras/stab126}, \href
  {https://ui.adsabs.harvard.edu/abs/2021MNRAS.502.2446E} {502, 2446}

\bibitem[\protect\citeauthoryear{{Erwin} et~al.,}{{Erwin}
  et~al.}{2023}]{Erwin_2023}
{Erwin} P.,  et~al., 2023, in prep

\bibitem[\protect\citeauthoryear{{Fathi}}{{Fathi}}{2004}]{Fathi_04}
{Fathi} K.,  2004, PhD thesis, University of Groningen

\bibitem[\protect\citeauthoryear{{Fathi}, {van de Ven}, {Peletier}, {Emsellem},
  {Falc{\'o}n-Barroso}, {Cappellari}  \& {de Zeeuw}}{{Fathi}
  et~al.}{2005}]{Fathi_05}
{Fathi} K.,  {van de Ven} G.,  {Peletier} R.~F.,  {Emsellem} E.,
  {Falc{\'o}n-Barroso} J.,  {Cappellari} M.,   {de Zeeuw} T.,  2005, \mn@doi
  [\mnras] {10.1111/j.1365-2966.2005.09648.x}, \href
  {https://ui.adsabs.harvard.edu/abs/2005MNRAS.364..773F} {364, 773}

\bibitem[\protect\citeauthoryear{{Fathi}, {Storchi-Bergmann}, {Riffel},
  {Winge}, {Axon}, {Robinson}, {Capetti}  \& {Marconi}}{{Fathi}
  et~al.}{2006}]{Fathi_06}
{Fathi} K.,  {Storchi-Bergmann} T.,  {Riffel} R.~A.,  {Winge} C.,  {Axon}
  D.~J.,  {Robinson} A.,  {Capetti} A.,   {Marconi} A.,  2006, \mn@doi [\apjl]
  {10.1086/503832}, \href
  {https://ui.adsabs.harvard.edu/abs/2006ApJ...641L..25F} {641, L25}

\bibitem[\protect\citeauthoryear{{Fathi} et~al.,}{{Fathi}
  et~al.}{2013}]{Fathi_13}
{Fathi} K.,  et~al., 2013, \mn@doi [\apjl] {10.1088/2041-8205/770/2/L27}, \href
  {https://ui.adsabs.harvard.edu/abs/2013ApJ...770L..27F} {770, L27}

\bibitem[\protect\citeauthoryear{{Fragkoudi}, {Athanassoula}  \&
  {Bosma}}{{Fragkoudi} et~al.}{2016}]{Fragkoudi_16}
{Fragkoudi} F.,  {Athanassoula} E.,   {Bosma} A.,  2016, \mn@doi [\mnras]
  {10.1093/mnrasl/slw120}, \href
  {https://ui.adsabs.harvard.edu/abs/2016MNRAS.462L..41F} {462, L41}

\bibitem[\protect\citeauthoryear{{Franx}, {van Gorkom}  \& {de Zeeuw}}{{Franx}
  et~al.}{1994}]{Franx_Gorkom_Zeeuw_1994}
{Franx} M.,  {van Gorkom} J.~H.,   {de Zeeuw} T.,  1994, \mn@doi [\apj]
  {10.1086/174939}, \href
  {https://ui.adsabs.harvard.edu/abs/1994ApJ...436..642F} {436, 642}

\bibitem[\protect\citeauthoryear{Freedman et~al.,}{Freedman
  et~al.}{2001}]{Freedman_2001}
Freedman W.~L.,  et~al., 2001, \mn@doi [The Astrophysical Journal]
  {10.1086/320638}, 553, 47

\bibitem[\protect\citeauthoryear{{Gadotti} et~al.,}{{Gadotti}
  et~al.}{2019}]{Gadotti_19}
{Gadotti} D.~A.,  et~al., 2019, \mn@doi [\mnras] {10.1093/mnras/sty2666}, \href
  {https://ui.adsabs.harvard.edu/abs/2019MNRAS.482..506G} {482, 506}

\bibitem[\protect\citeauthoryear{{Gadotti} et~al.,}{{Gadotti}
  et~al.}{2020}]{Gadotti_20}
{Gadotti} D.~A.,  et~al., 2020, \mn@doi [\aap] {10.1051/0004-6361/202038448},
  \href {https://ui.adsabs.harvard.edu/abs/2020A&A...643A..14G} {643, A14}

\bibitem[\protect\citeauthoryear{{Garc{\'\i}a-Burillo}, {Combes}, {Schinnerer},
  {Boone}  \& {Hunt}}{{Garc{\'\i}a-Burillo} et~al.}{2005}]{Garcia-Burillo-05}
{Garc{\'\i}a-Burillo} S.,  {Combes} F.,  {Schinnerer} E.,  {Boone} F.,   {Hunt}
  L.~K.,  2005, \mn@doi [\aap] {10.1051/0004-6361:20052900}, \href
  {https://ui.adsabs.harvard.edu/abs/2005A&A...441.1011G} {441, 1011}

\bibitem[\protect\citeauthoryear{{Garc{\'\i}a-Burillo}
  et~al.,}{{Garc{\'\i}a-Burillo} et~al.}{2014}]{Garcia_Burillo_2014}
{Garc{\'\i}a-Burillo} S.,  et~al., 2014, \mn@doi [\aap]
  {10.1051/0004-6361/201423843}, \href
  {https://ui.adsabs.harvard.edu/abs/2014A&A...567A.125G} {567, A125}

\bibitem[\protect\citeauthoryear{{Heller} \& {Shlosman}}{{Heller} \&
  {Shlosman}}{1994}]{Heller_1994}
{Heller} C.~H.,  {Shlosman} I.,  1994, \mn@doi [\apj] {10.1086/173874}, \href
  {https://ui.adsabs.harvard.edu/abs/1994ApJ...424...84H} {424, 84}

\bibitem[\protect\citeauthoryear{{Hermosa Mu{\~n}oz}, {M{\'a}rquez}, {Cazzoli},
  {Masegosa}  \& {Ag{\'\i}s-Gonz{\'a}lez}}{{Hermosa Mu{\~n}oz}
  et~al.}{2022}]{Hermosa_Munoz_22}
{Hermosa Mu{\~n}oz} L.,  {M{\'a}rquez} I.,  {Cazzoli} S.,  {Masegosa} J.,
  {Ag{\'\i}s-Gonz{\'a}lez} B.,  2022, \mn@doi [\aap]
  {10.1051/0004-6361/202142629}, \href
  {https://ui.adsabs.harvard.edu/abs/2022A&A...660A.133H} {660, A133}

\bibitem[\protect\citeauthoryear{{Ho}}{{Ho}}{2008}]{Ho_08}
{Ho} L.~C.,  2008, \mn@doi [\araa] {10.1146/annurev.astro.45.051806.110546},
  \href {https://ui.adsabs.harvard.edu/abs/2008ARA&A..46..475H} {46, 475}

\bibitem[\protect\citeauthoryear{{Hu}, {Lazarian}, {Beck}  \& {Xu}}{{Hu}
  et~al.}{2022}]{Hu_22}
{Hu} Y.,  {Lazarian} A.,  {Beck} R.,   {Xu} S.,  2022, arXiv e-prints, \href
  {https://ui.adsabs.harvard.edu/abs/2022arXiv220605423H} {p. arXiv:2206.05423}

\bibitem[\protect\citeauthoryear{{Izumi} et~al.,}{{Izumi}
  et~al.}{2013}]{Izumi_13}
{Izumi} T.,  et~al., 2013, \mn@doi [\pasj] {10.1093/pasj/65.5.100}, \href
  {https://ui.adsabs.harvard.edu/abs/2013PASJ...65..100I} {65, 100}

\bibitem[\protect\citeauthoryear{{Izumi} et~al.,}{{Izumi}
  et~al.}{2017}]{Izumi_17}
{Izumi} T.,  et~al., 2017, \mn@doi [\apjl] {10.3847/2041-8213/aa808f}, \href
  {https://ui.adsabs.harvard.edu/abs/2017ApJ...845L...5I} {845, L5}

\bibitem[\protect\citeauthoryear{{Kauffmann} et~al.,}{{Kauffmann}
  et~al.}{2003}]{Kauffmann_03}
{Kauffmann} G.,  et~al., 2003, \mn@doi [\mnras]
  {10.1111/j.1365-2966.2003.07154.x}, \href
  {https://ui.adsabs.harvard.edu/abs/2003MNRAS.346.1055K} {346, 1055}

\bibitem[\protect\citeauthoryear{{Kewley}, {Dopita}, {Sutherland}, {Heisler}
  \& {Trevena}}{{Kewley} et~al.}{2001}]{Kewley_01a}
{Kewley} L.~J.,  {Dopita} M.~A.,  {Sutherland} R.~S.,  {Heisler} C.~A.,
  {Trevena} J.,  2001, \mn@doi [\apj] {10.1086/321545}, \href
  {https://ui.adsabs.harvard.edu/abs/2001ApJ...556..121K} {556, 121}

\bibitem[\protect\citeauthoryear{{Kewley}, {Groves}, {Kauffmann}  \&
  {Heckman}}{{Kewley} et~al.}{2006}]{Kewley_06}
{Kewley} L.~J.,  {Groves} B.,  {Kauffmann} G.,   {Heckman} T.,  2006, \mn@doi
  [\mnras] {10.1111/j.1365-2966.2006.10859.x}, \href
  {https://ui.adsabs.harvard.edu/abs/2006MNRAS.372..961K} {372, 961}

\bibitem[\protect\citeauthoryear{{Kim}, {Seo}, {Stone}, {Yoon}  \&
  {Teuben}}{{Kim} et~al.}{2012}]{Kim_2012}
{Kim} W.-T.,  {Seo} W.-Y.,  {Stone} J.~M.,  {Yoon} D.,   {Teuben} P.~J.,  2012,
  \mn@doi [\apj] {10.1088/0004-637X/747/1/60}, \href
  {https://ui.adsabs.harvard.edu/abs/2012ApJ...747...60K} {747, 60}

\bibitem[\protect\citeauthoryear{{Krajnovi{\'c}}, {Cappellari}, {de Zeeuw}  \&
  {Copin}}{{Krajnovi{\'c}} et~al.}{2006}]{Krajnovic_06}
{Krajnovi{\'c}} D.,  {Cappellari} M.,  {de Zeeuw} P.~T.,   {Copin} Y.,  2006,
  \mn@doi [\mnras] {10.1111/j.1365-2966.2005.09902.x}, \href
  {https://ui.adsabs.harvard.edu/abs/2006MNRAS.366..787K} {366, 787}

\bibitem[\protect\citeauthoryear{{Lang} et~al.,}{{Lang} et~al.}{2020}]{Lang_20}
{Lang} P.,  et~al., 2020, \mn@doi [\apj] {10.3847/1538-4357/ab9953}, \href
  {https://ui.adsabs.harvard.edu/abs/2020ApJ...897..122L} {897, 122}

\bibitem[\protect\citeauthoryear{{Legodi}, {Taylor}  \& {Stil}}{{Legodi}
  et~al.}{2021}]{Legodi_21}
{Legodi} L.~S.,  {Taylor} A.~R.,   {Stil} J.~M.,  2021, \mn@doi [\mnras]
  {10.1093/mnras/staa3266}, \href
  {https://ui.adsabs.harvard.edu/abs/2021MNRAS.500..576L} {500, 576}

\bibitem[\protect\citeauthoryear{Leroy et~al.,}{Leroy
  et~al.}{2021}]{Leroy_2021b}
Leroy A.~K.,  et~al., 2021, \mn@doi [The Astrophysical Journal Supplement
  Series] {10.3847/1538-4365/ac17f3}, 257, 43

\bibitem[\protect\citeauthoryear{{Lin}, {Wang}, {Hsieh}, {Taam}, {Yang}  \&
  {Yen}}{{Lin} et~al.}{2013}]{Lin_13}
{Lin} L.-H.,  {Wang} H.-H.,  {Hsieh} P.-Y.,  {Taam} R.~E.,  {Yang} C.-C.,
  {Yen} D. C.~C.,  2013, \mn@doi [\apj] {10.1088/0004-637X/771/1/8}, \href
  {https://ui.adsabs.harvard.edu/abs/2013ApJ...771....8L} {771, 8}

\bibitem[\protect\citeauthoryear{{Lopez-Rodriguez} et~al.,}{{Lopez-Rodriguez}
  et~al.}{2021}]{Lopez-Rodriguez_21}
{Lopez-Rodriguez} E.,  et~al., 2021, \mn@doi [\apj] {10.3847/1538-4357/ac2e01},
  \href {https://ui.adsabs.harvard.edu/abs/2021ApJ...923..150L} {923, 150}

\bibitem[\protect\citeauthoryear{{Maciejewski}}{{Maciejewski}}{2004}]{Maciejewski_2004}
{Maciejewski} W.,  2004, \mn@doi [\mnras] {10.1111/j.1365-2966.2004.08254.x},
  \href {https://ui.adsabs.harvard.edu/abs/2004MNRAS.354..892M} {354, 892}

\bibitem[\protect\citeauthoryear{{Maciejewski}, {Teuben}, {Sparke}  \&
  {Stone}}{{Maciejewski} et~al.}{2002}]{Maciejewski_02}
{Maciejewski} W.,  {Teuben} P.~J.,  {Sparke} L.~S.,   {Stone} J.~M.,  2002,
  \mn@doi [\mnras] {10.1046/j.1365-8711.2002.04957.x}, \href
  {https://ui.adsabs.harvard.edu/abs/2002MNRAS.329..502M} {329, 502}

\bibitem[\protect\citeauthoryear{{Man}, {Peng}, {Kong}, {Guo}, {Zhang}  \&
  {Dou}}{{Man} et~al.}{2019}]{Man_19}
{Man} Z.-y.,  {Peng} Y.-j.,  {Kong} X.,  {Guo} K.-x.,  {Zhang} C.-p.,   {Dou}
  J.,  2019, \mn@doi [\mnras] {10.1093/mnras/stz1706}, \href
  {https://ui.adsabs.harvard.edu/abs/2019MNRAS.488...89M} {488, 89}

\bibitem[\protect\citeauthoryear{{M{\'a}rquez}, {Masegosa},
  {Gonz{\'a}lez-Martin}, {Hern{\'a}ndez-Garcia}, {Povi{\'c}}, {Netzer},
  {Cazzoli}  \& {del Olmo}}{{M{\'a}rquez} et~al.}{2017}]{Marquez_17}
{M{\'a}rquez} I.,  {Masegosa} J.,  {Gonz{\'a}lez-Martin} O.,
  {Hern{\'a}ndez-Garcia} L.,  {Povi{\'c}} M.,  {Netzer} H.,  {Cazzoli} S.,
  {del Olmo} A.,  2017, \mn@doi [Frontiers in Astronomy and Space Sciences]
  {10.3389/fspas.2017.00034}, \href
  {https://ui.adsabs.harvard.edu/abs/2017FrASS...4...34M} {4, 34}

\bibitem[\protect\citeauthoryear{{Mart{\'\i}n} et~al.,}{{Mart{\'\i}n}
  et~al.}{2015}]{Martin_15}
{Mart{\'\i}n} S.,  et~al., 2015, \mn@doi [\aap] {10.1051/0004-6361/201425105},
  \href {https://ui.adsabs.harvard.edu/abs/2015A&A...573A.116M} {573, A116}

\bibitem[\protect\citeauthoryear{{Martini}, {Regan}, {Mulchaey}  \&
  {Pogge}}{{Martini} et~al.}{2003}]{Martini_03}
{Martini} P.,  {Regan} M.~W.,  {Mulchaey} J.~S.,   {Pogge} R.~W.,  2003,
  \mn@doi [\apj] {10.1086/374685}, \href
  {https://ui.adsabs.harvard.edu/abs/2003ApJ...589..774M} {589, 774}

\bibitem[\protect\citeauthoryear{{Mazzalay} et~al.,}{{Mazzalay}
  et~al.}{2014}]{Mazzalay_14}
{Mazzalay} X.,  et~al., 2014, \mn@doi [\mnras] {10.1093/mnras/stt2319}, \href
  {https://ui.adsabs.harvard.edu/abs/2014MNRAS.438.2036M} {438, 2036}

\bibitem[\protect\citeauthoryear{{Mazzuca}, {Knapen}, {Veilleux}  \&
  {Regan}}{{Mazzuca} et~al.}{2008}]{Mazzuca_08}
{Mazzuca} L.~M.,  {Knapen} J.~H.,  {Veilleux} S.,   {Regan} M.~W.,  2008,
  \mn@doi [\apjs] {10.1086/522338}, \href
  {https://ui.adsabs.harvard.edu/abs/2008ApJS..174..337M} {174, 337}

\bibitem[\protect\citeauthoryear{{Nemmen}, {Storchi-Bergmann}, {Yuan},
  {Eracleous}, {Terashima}  \& {Wilson}}{{Nemmen} et~al.}{2006}]{Nemmen_06}
{Nemmen} R.~S.,  {Storchi-Bergmann} T.,  {Yuan} F.,  {Eracleous} M.,
  {Terashima} Y.,   {Wilson} A.~S.,  2006, \mn@doi [\apj] {10.1086/500571},
  \href {https://ui.adsabs.harvard.edu/abs/2006ApJ...643..652N} {643, 652}

\bibitem[\protect\citeauthoryear{{Osterbrock} \& {Ferland}}{{Osterbrock} \&
  {Ferland}}{2006}]{Osterbrock_book}
{Osterbrock} D.~E.,  {Ferland} G.~J.,  2006, {Astrophysics of gaseous nebulae
  and active galactic nuclei}.
University Science Books

\bibitem[\protect\citeauthoryear{{Prieto}, {Maciejewski}  \&
  {Reunanen}}{{Prieto} et~al.}{2005}]{Prieto_05}
{Prieto} M.~A.,  {Maciejewski} W.,   {Reunanen} J.,  2005, \mn@doi [\aj]
  {10.1086/444591}, \href
  {https://ui.adsabs.harvard.edu/abs/2005AJ....130.1472P} {130, 1472}

\bibitem[\protect\citeauthoryear{{Prieto}, {Fernandez-Ontiveros}, {Bruzual},
  {Burkert}, {Schartmann}  \& {Charlot}}{{Prieto} et~al.}{2019}]{Prieto_19}
{Prieto} M.~A.,  {Fernandez-Ontiveros} J.~A.,  {Bruzual} G.,  {Burkert} A.,
  {Schartmann} M.,   {Charlot} S.,  2019, \mn@doi [\mnras]
  {10.1093/mnras/stz579}, \href
  {https://ui.adsabs.harvard.edu/abs/2019MNRAS.485.3264P} {485, 3264}

\bibitem[\protect\citeauthoryear{{Quillen}, {Frogel}, {Kuchinski}  \&
  {Terndrup}}{{Quillen} et~al.}{1995}]{Quillen_95}
{Quillen} A.~C.,  {Frogel} J.~A.,  {Kuchinski} L.~E.,   {Terndrup} D.~M.,
  1995, \mn@doi [\aj] {10.1086/117503}, \href
  {https://ui.adsabs.harvard.edu/abs/1995AJ....110..156Q} {110, 156}

\bibitem[\protect\citeauthoryear{{Rogstad}, {Lockhart}  \& {Wright}}{{Rogstad}
  et~al.}{1974}]{Rogstad_1974}
{Rogstad} D.~H.,  {Lockhart} I.~A.,   {Wright} M.~C.~H.,  1974, \mn@doi [\apj]
  {10.1086/153164}, \href
  {https://ui.adsabs.harvard.edu/abs/1974ApJ...193..309R} {193, 309}

\bibitem[\protect\citeauthoryear{{Sanders} \& {Huntley}}{{Sanders} \&
  {Huntley}}{1976}]{Sanders_1976}
{Sanders} R.~H.,  {Huntley} J.~M.,  1976, \mn@doi [\apj] {10.1086/154692},
  \href {https://ui.adsabs.harvard.edu/abs/1976ApJ...209...53S} {209, 53}

\bibitem[\protect\citeauthoryear{{Santoro}, {Tadhunter}, {Baron}, {Morganti}
  \& {Holt}}{{Santoro} et~al.}{2020}]{Santoro_20}
{Santoro} F.,  {Tadhunter} C.,  {Baron} D.,  {Morganti} R.,   {Holt} J.,  2020,
  \mn@doi [\aap] {10.1051/0004-6361/202039077}, \href
  {https://ui.adsabs.harvard.edu/abs/2020A&A...644A..54S} {644, A54}

\bibitem[\protect\citeauthoryear{{Sarzi} et~al.,}{{Sarzi}
  et~al.}{2006}]{Sarzi_06}
{Sarzi} M.,  et~al., 2006, \mn@doi [\mnras] {10.1111/j.1365-2966.2005.09839.x},
  \href {https://ui.adsabs.harvard.edu/abs/2006MNRAS.366.1151S} {366, 1151}

\bibitem[\protect\citeauthoryear{{Schoenmakers}, {Franx}  \& {de
  Zeeuw}}{{Schoenmakers} et~al.}{1997}]{Schoenmakers_1997}
{Schoenmakers} R.~H.~M.,  {Franx} M.,   {de Zeeuw} P.~T.,  1997, \mn@doi
  [\mnras] {10.1093/mnras/292.2.349}, \href
  {https://ui.adsabs.harvard.edu/abs/1997MNRAS.292..349S} {292, 349}

\bibitem[\protect\citeauthoryear{{Shaw}, {Combes}, {Axon}  \& {Wright}}{{Shaw}
  et~al.}{1993}]{Shaw_1993}
{Shaw} M.~A.,  {Combes} F.,  {Axon} D.~J.,   {Wright} G.~S.,  1993, \aap, \href
  {https://ui.adsabs.harvard.edu/abs/1993A&A...273...31S} {273, 31}

\bibitem[\protect\citeauthoryear{{Shlosman}}{{Shlosman}}{2001}]{Shlosman_01}
{Shlosman} I.,  2001, in {Knapen} J.~H.,  {Beckman} J.~E.,  {Shlosman} I.,
  {Mahoney} T.~J.,  eds,  Astronomical Society of the Pacific Conference Series
  Vol. 249, The Central Kiloparsec of Starbursts and AGN: The La Palma
  Connection. p.~55

\bibitem[\protect\citeauthoryear{{Shlosman}, {Frank}  \& {Begelman}}{{Shlosman}
  et~al.}{1989}]{Shlosman_1989}
{Shlosman} I.,  {Frank} J.,   {Begelman} M.~C.,  1989, \mn@doi [\nat]
  {10.1038/338045a0}, \href
  {https://ui.adsabs.harvard.edu/abs/1989Natur.338...45S} {338, 45}

\bibitem[\protect\citeauthoryear{{Sormani}, {Binney}  \& {Magorrian}}{{Sormani}
  et~al.}{2015}]{Sormani_15}
{Sormani} M.~C.,  {Binney} J.,   {Magorrian} J.,  2015, \mn@doi [\mnras]
  {10.1093/mnras/stv441}, \href
  {https://ui.adsabs.harvard.edu/abs/2015MNRAS.449.2421S} {449, 2421}

\bibitem[\protect\citeauthoryear{{Spekkens} \& {Sellwood}}{{Spekkens} \&
  {Sellwood}}{2007}]{Spekkens_Sellwood_07}
{Spekkens} K.,  {Sellwood} J.~A.,  2007, \mn@doi [\apj] {10.1086/518471}, \href
  {https://ui.adsabs.harvard.edu/abs/2007ApJ...664..204S} {664, 204}

\bibitem[\protect\citeauthoryear{{Storchi-Bergmann}, {Rodriguez-Ardila},
  {Schmitt}, {Wilson}  \& {Baldwin}}{{Storchi-Bergmann}
  et~al.}{1996}]{Storchi-Bergmann_1996}
{Storchi-Bergmann} T.,  {Rodriguez-Ardila} A.,  {Schmitt} H.~R.,  {Wilson}
  A.~S.,   {Baldwin} J.~A.,  1996, \mn@doi [\apj] {10.1086/178043}, \href
  {https://ui.adsabs.harvard.edu/abs/1996ApJ...472...83S} {472, 83}

\bibitem[\protect\citeauthoryear{{Storchi-Bergmann} et~al.,}{{Storchi-Bergmann}
  et~al.}{2003}]{Storchi-Bergmann_03}
{Storchi-Bergmann} T.,  et~al., 2003, \mn@doi [\apj] {10.1086/378938}, \href
  {https://ui.adsabs.harvard.edu/abs/2003ApJ...598..956S} {598, 956}

\bibitem[\protect\citeauthoryear{{Storchi-Bergmann}, {Dors}, {Riffel}, {Fathi},
  {Axon}, {Robinson}, {Marconi}  \& {{\"O}stlin}}{{Storchi-Bergmann}
  et~al.}{2007}]{Storchi-Bergmann-07}
{Storchi-Bergmann} T.,  {Dors} Oli~L. J.,  {Riffel} R.~A.,  {Fathi} K.,  {Axon}
  D.~J.,  {Robinson} A.,  {Marconi} A.,   {{\"O}stlin} G.,  2007, \mn@doi
  [\apj] {10.1086/521918}, \href
  {https://ui.adsabs.harvard.edu/abs/2007ApJ...670..959S} {670, 959}

\bibitem[\protect\citeauthoryear{{Tody}}{{Tody}}{2000}]{IRAF_2000}
{Tody} D.,  2000, in {Murdin} P.,  ed., , Encyclopedia of Astronomy and
  Astrophysics.
p.~2923, \mn@doi{10.1888/0333750888/2923}

\bibitem[\protect\citeauthoryear{{Vazdekis} et~al.,}{{Vazdekis}
  et~al.}{2015}]{Vazdekis_15}
{Vazdekis} A.,  et~al., 2015, \mn@doi [\mnras] {10.1093/mnras/stv151}, \href
  {https://ui.adsabs.harvard.edu/abs/2015MNRAS.449.1177V} {449, 1177}

\bibitem[\protect\citeauthoryear{{Zurita}, {Rela{\~n}o}, {Beckman}  \&
  {Knapen}}{{Zurita} et~al.}{2004}]{Zurita_04}
{Zurita} A.,  {Rela{\~n}o} M.,  {Beckman} J.~E.,   {Knapen} J.~H.,  2004,
  \mn@doi [\aap] {10.1051/0004-6361:20031049}, \href
  {https://ui.adsabs.harvard.edu/abs/2004A&A...413...73Z} {413, 73}

\bibitem[\protect\citeauthoryear{{van de Ven} \& {Fathi}}{{van de Ven} \&
  {Fathi}}{2010}]{vandeVen_Fathi_2010}
{van de Ven} G.,  {Fathi} K.,  2010, \mn@doi [\apj]
  {10.1088/0004-637X/723/1/767}, \href
  {https://ui.adsabs.harvard.edu/abs/2010ApJ...723..767V} {723, 767}

\bibitem[\protect\citeauthoryear{{van der Kruit} \& {Allen}}{{van der Kruit} \&
  {Allen}}{1978}]{van_der_Kruit-Allen-1978}
{van der Kruit} P.~C.,  {Allen} R.~J.,  1978, \mn@doi [\araa]
  {10.1146/annurev.aa.16.090178.000535}, \href
  {https://ui.adsabs.harvard.edu/abs/1978ARA&A..16..103V} {16, 103}

\bibitem[\protect\citeauthoryear{{van der Marel} \& {Franx}}{{van der Marel} \&
  {Franx}}{1993}]{van_der_Marel_1993}
{van der Marel} R.~P.,  {Franx} M.,  1993, \mn@doi [\apj] {10.1086/172534},
  \href {https://ui.adsabs.harvard.edu/abs/1993ApJ...407..525V} {407, 525}

\makeatother
\end{thebibliography}
\appendix
\counterwithin{figure}{section}

\section{Supplementary Maps}
\label{app_supplementary_maps}
\begin{figure}
    \begin{subfigure}[ ]{0.25\textwidth}
    \includegraphics[width=1\textwidth]{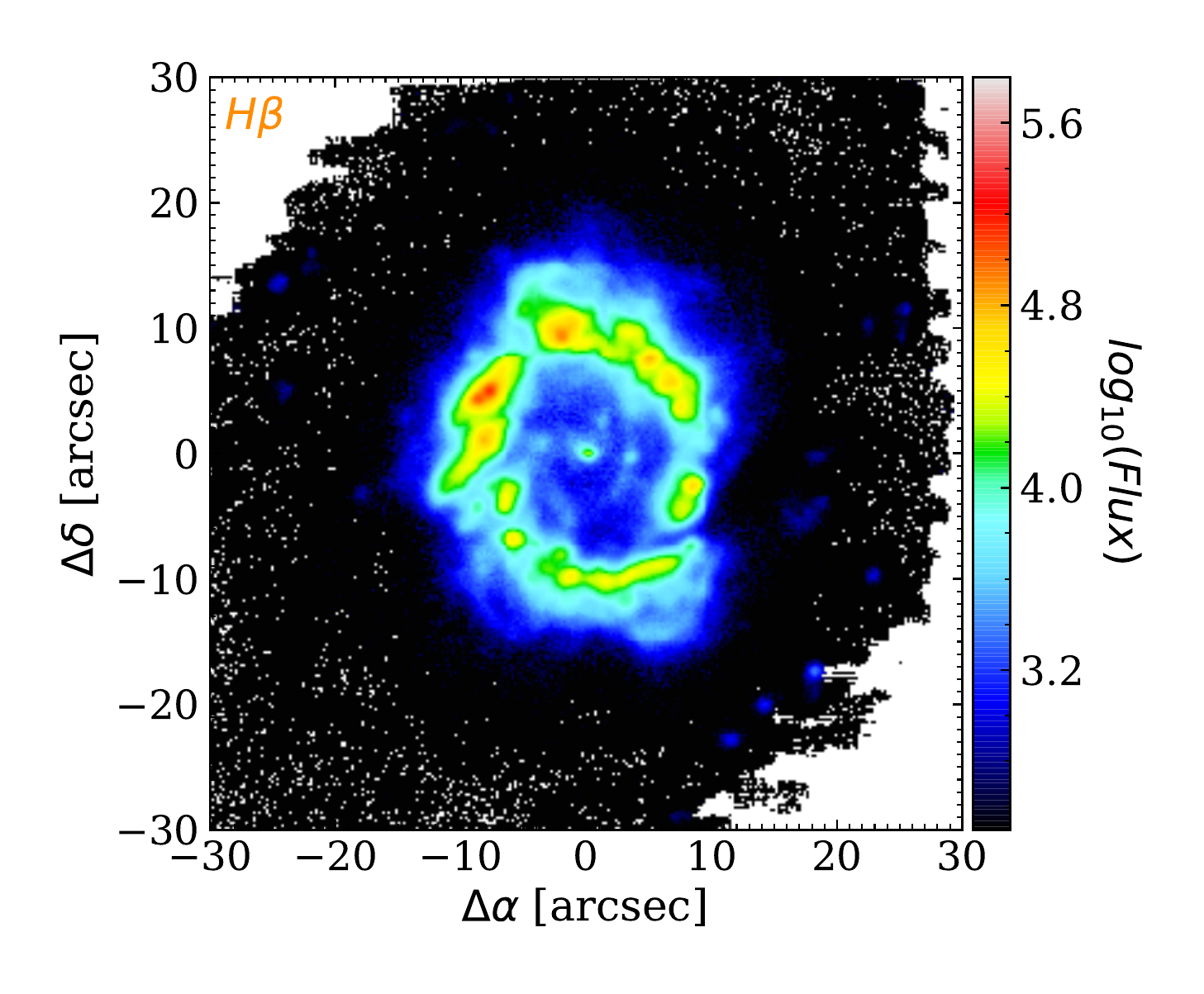}
    \end{subfigure}
    \hspace{-0.6cm}
    \begin{subfigure}[ ]{0.25\textwidth}
    \includegraphics[width=1\textwidth]{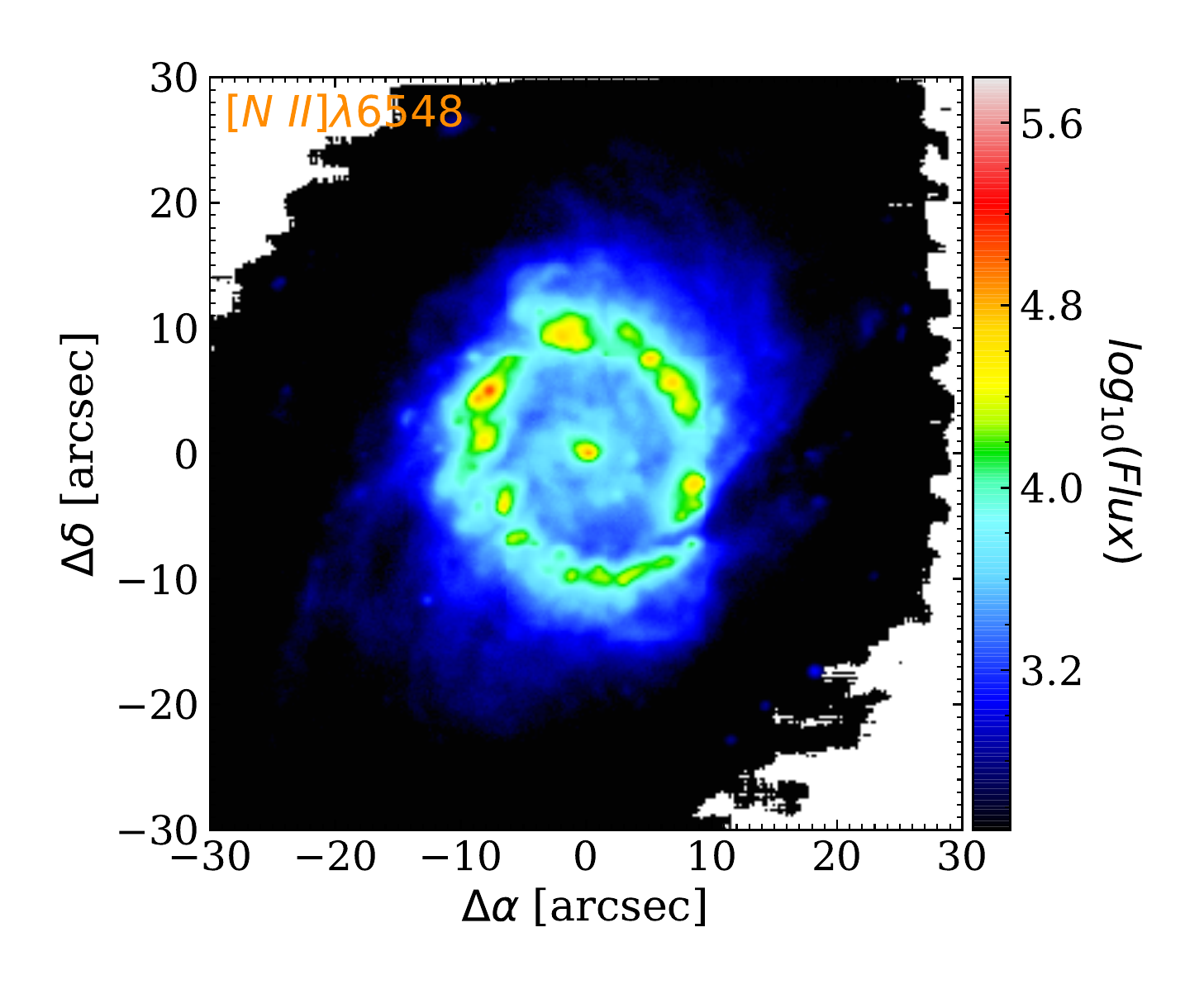}
    \end{subfigure}
    \begin{subfigure}[ ]{0.25\textwidth}
    \includegraphics[width=1\textwidth]{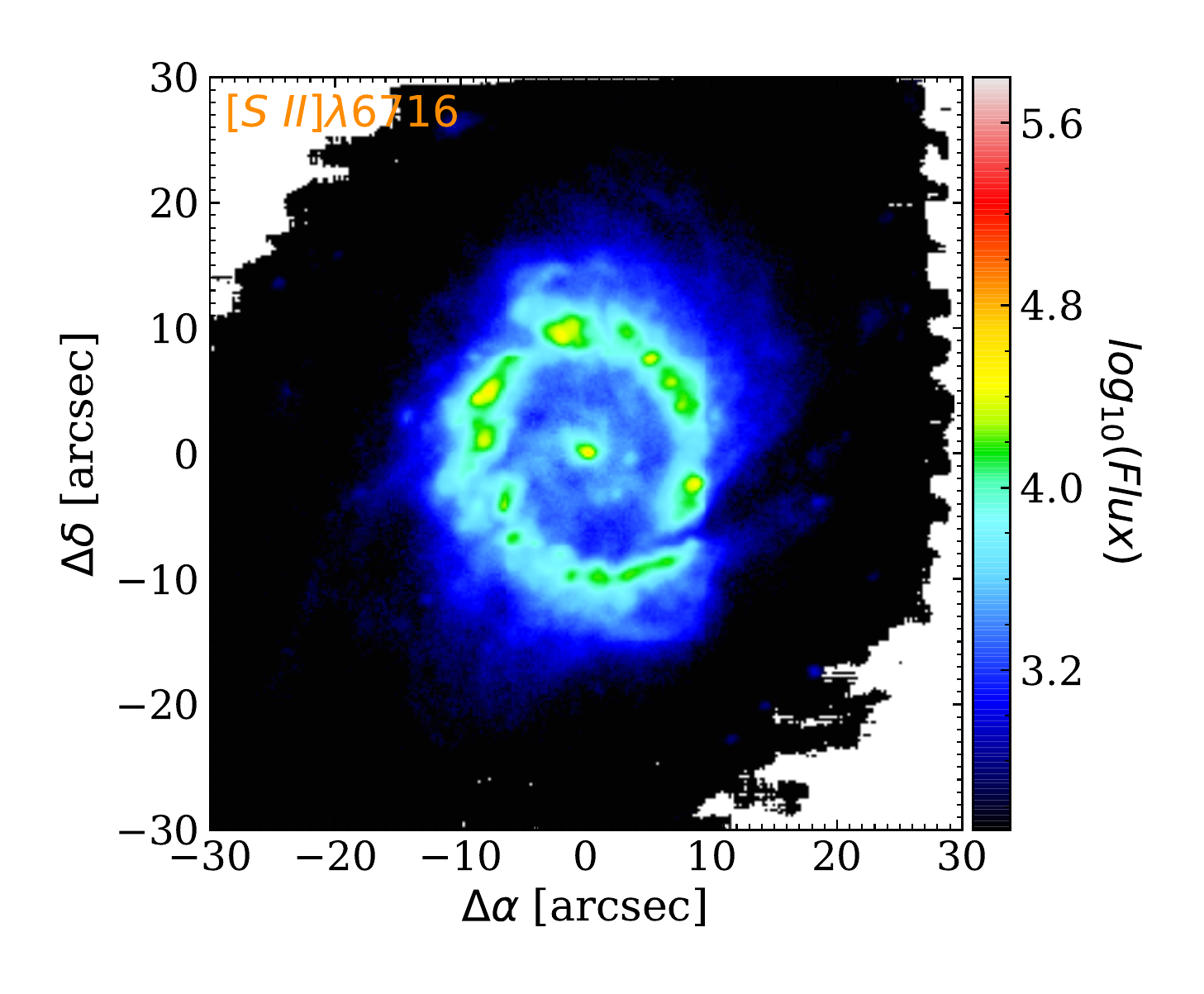}
    \end{subfigure} 
    \hspace{-0.6cm}
    \begin{subfigure}[ ]{0.25\textwidth}
    \includegraphics[width=1\textwidth]{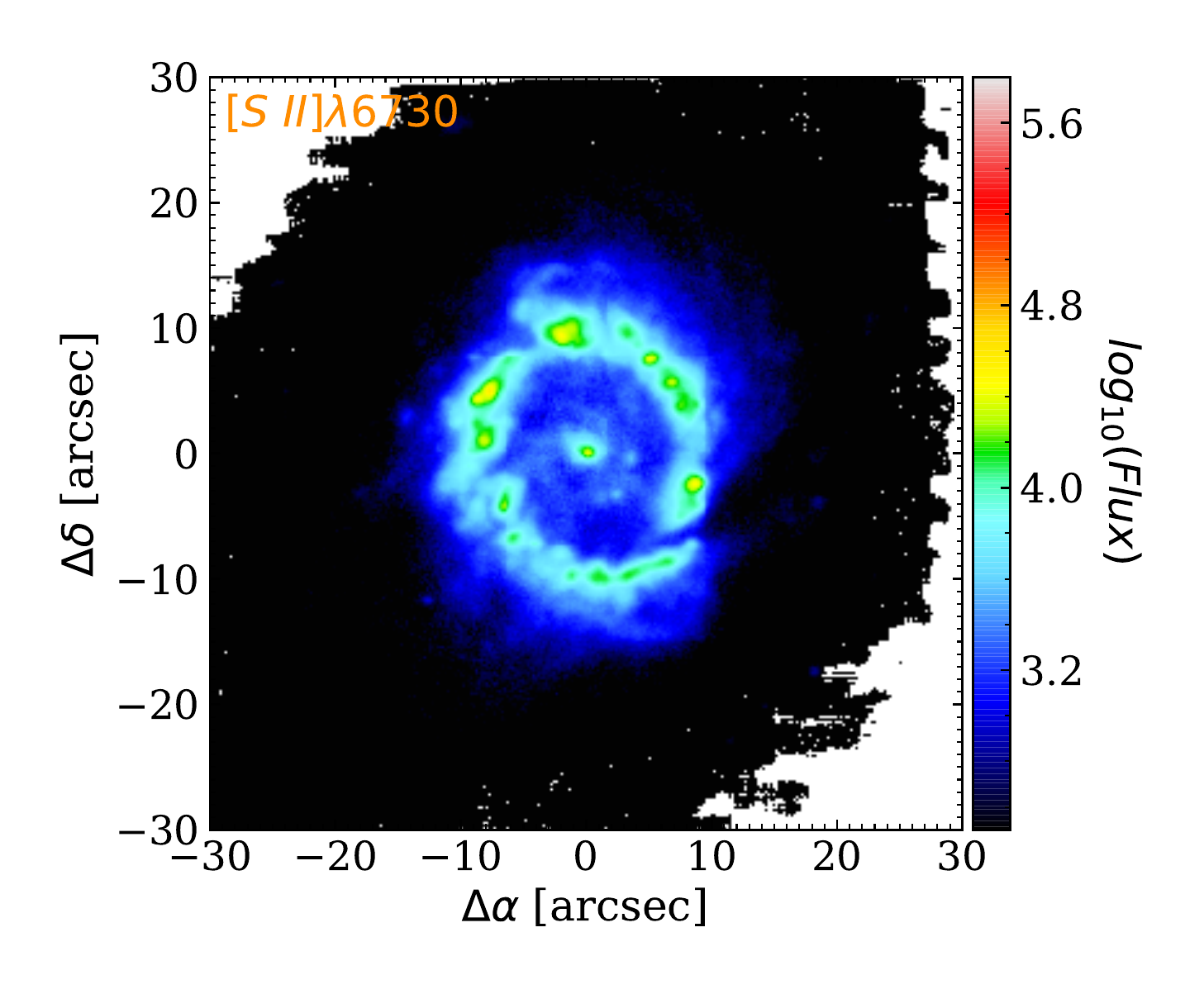}
    \end{subfigure}
\caption{Flux maps of the weaker lines in each emission line group outlined in Table \ref{tab:tracers}. Units are in $10^{-20}$ergs$^{-1}$cm$^{-2}$spaxel$^{-1}$ units.}
\label{appfig:others_flux_subplots}
\end{figure}

\begin{figure}
    \hspace{-0.45cm}
    \begin{subfigure}[ ]{0.25\textwidth}
    \includegraphics[width=\textwidth]{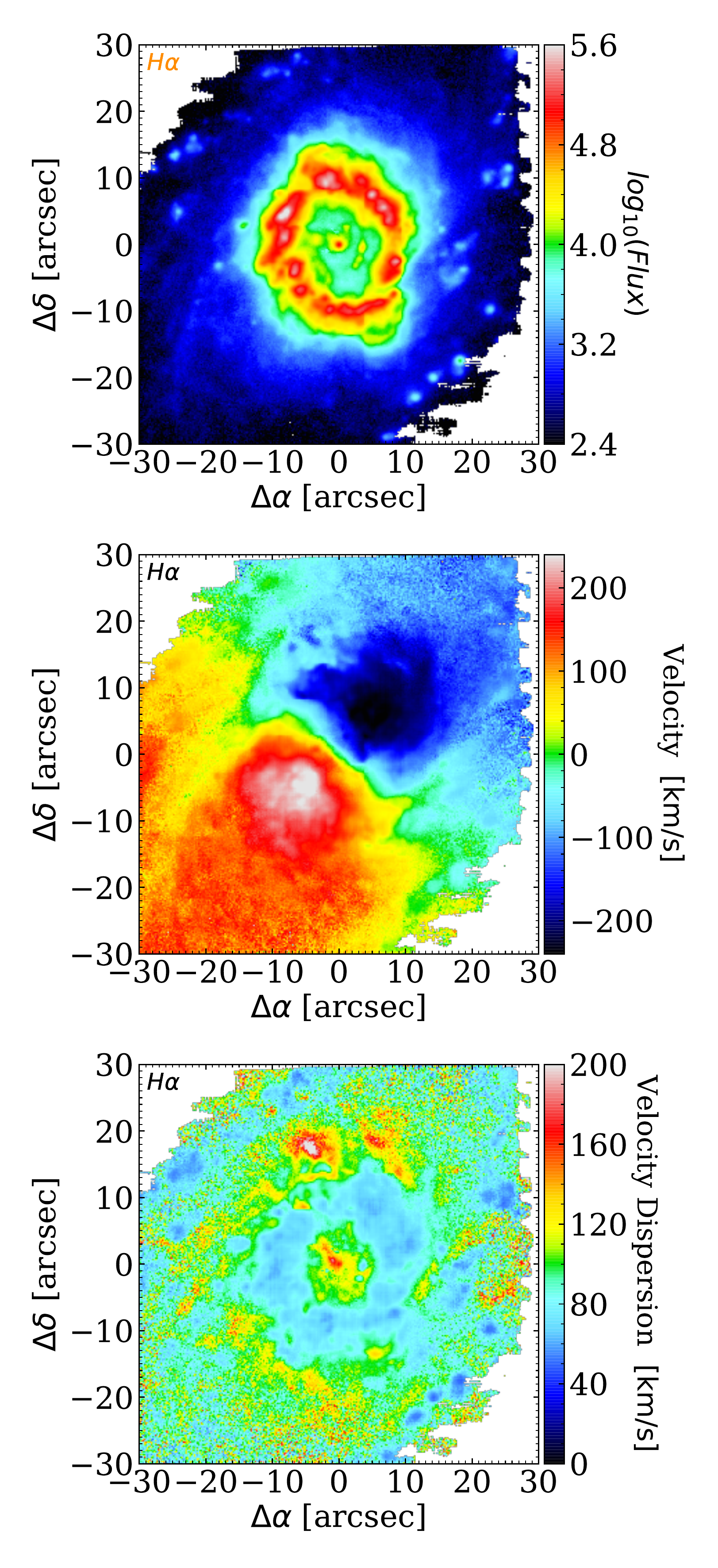}
    \end{subfigure}
    \begin{subfigure}[ ]{0.25\textwidth}
    \includegraphics[width=\textwidth]{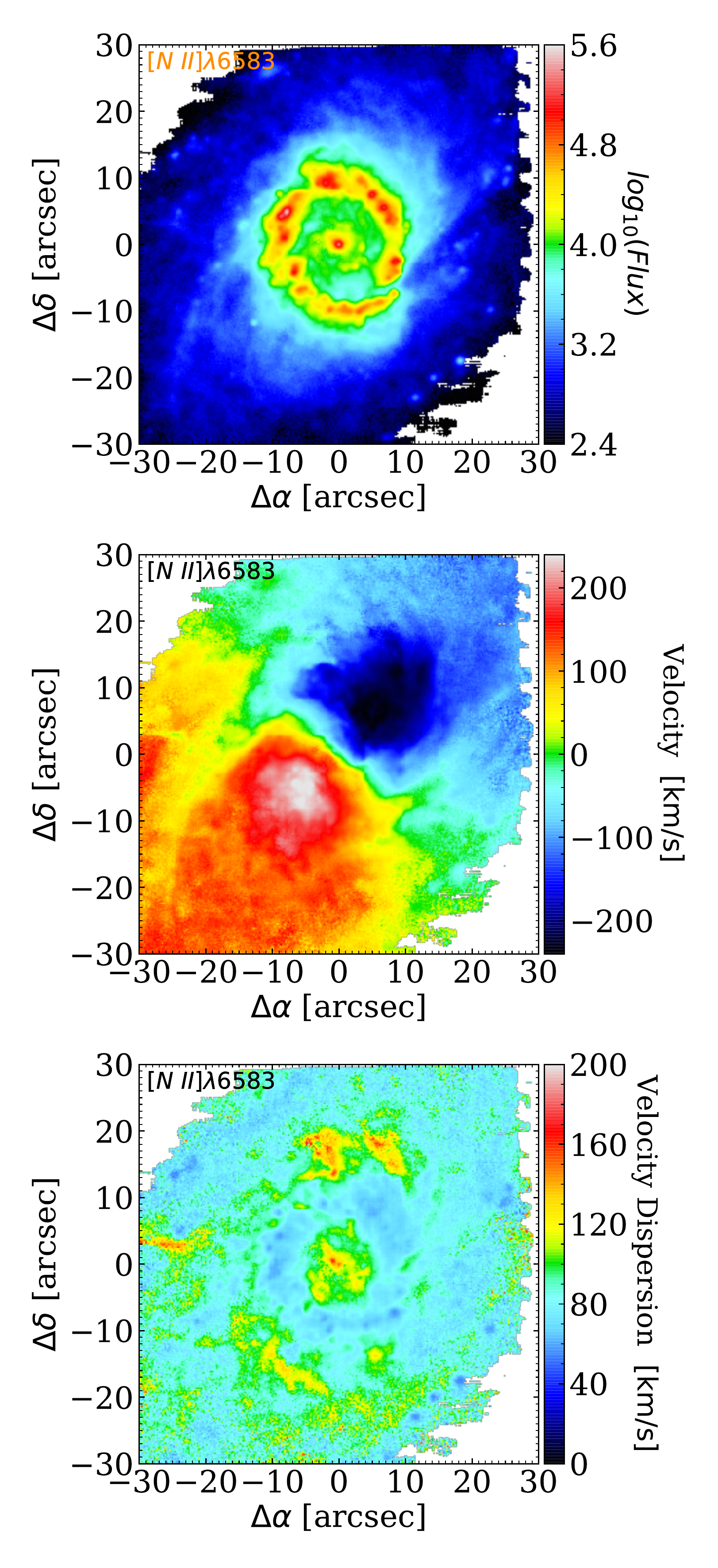}
    \end{subfigure}
 \caption{Corrected flux and LOSVD moments after combining the results from single Gaussian fits with the fits including the secondary component (as explained in Sect. \ref{sec:excess}). For \ha~line (left), and for \nii~line (right). We display the line flux on a logarithmic scale
in units $10^{-20}$ergs$^{-1}$cm$^{-2}$spaxel$^{-1}$ (top row), velocity (middle row) and velocity dispersion (bottom row).}
\label{fig:HA_NII_gaskinematics_corrected}
\end{figure}

\begin{figure*}
    \includegraphics[width=1\textwidth]{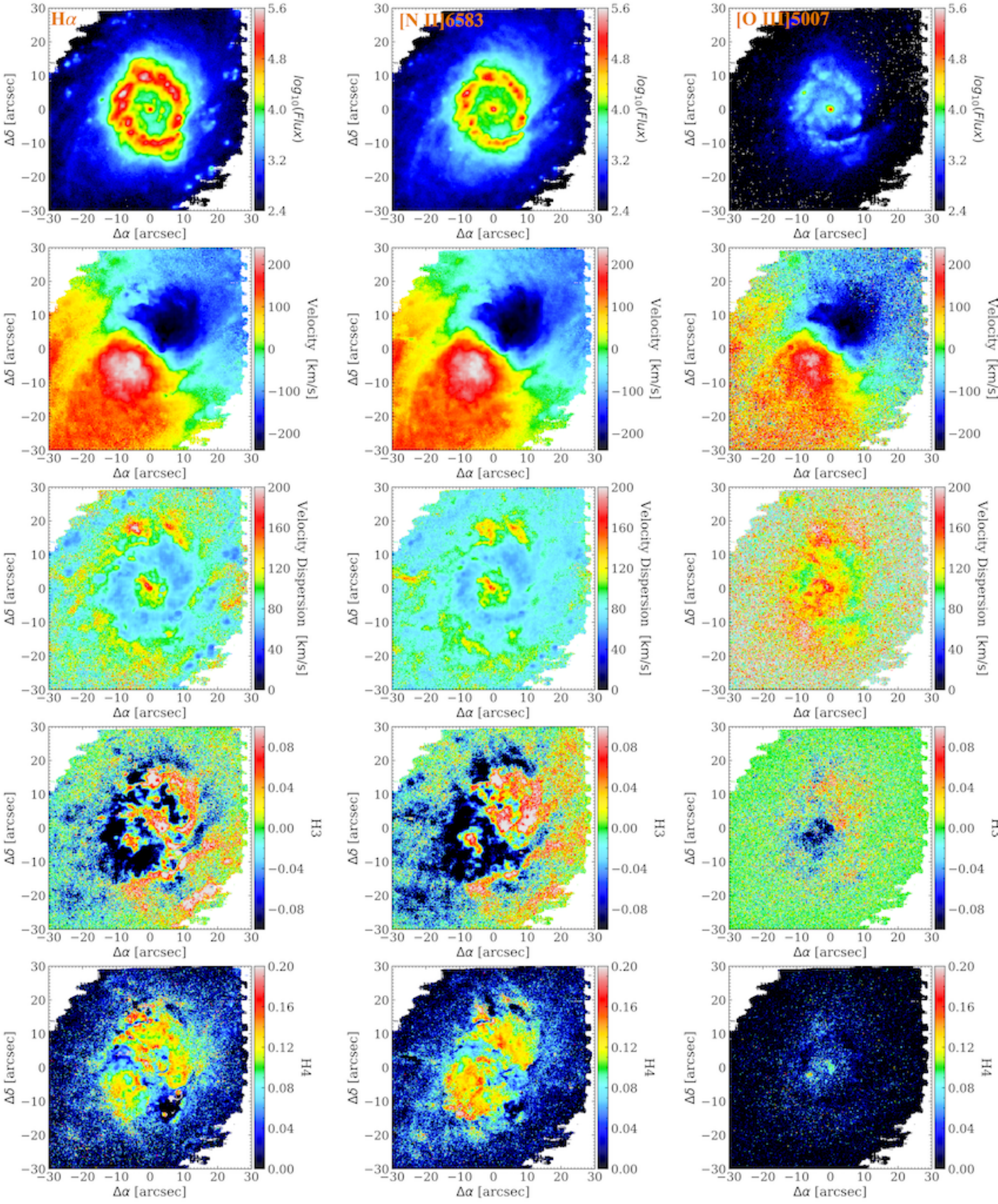}
    \caption{Flux and LOSVD moments for fits including higher order moments (h3 and h4). Columns show \ha~(left), \nii~(middle) and \oiii~(right). The units of the flux measurements are $10^{-20}$ergs$^{-1}$cm$^{-2}$spaxel$^{-1}$.}
    \label{fig:NGC1097_DAP_h3h4}
\end{figure*}

\end{document}